\documentclass[aps,amsmath,amssymb,showpacs,showkeys]{revtex4}
\usepackage[dvips]{graphicx,color}
\usepackage{times}
\usepackage{xcolor}
\usepackage[%
  colorlinks=true,
  urlcolor=blue,
  linkcolor=red,
  citecolor=blue
]{hyperref}

\newcommand{\orcid}[1]{\href{https://orcid.org/#1}{\includegraphics[width=8pt]{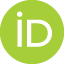}}}

\begin{document}
\title{Cosmology with a new $f(R)$ gravity model in Palatini formalism}

\author{Dhruba Jyoti Gogoi \orcid{0000-0002-4776-8506}}
\email[Email: ]{moloydhruba@yahoo.in}

\affiliation{Department of Physics, Dibrugarh University,
Dibrugarh 786004, Assam, India}

\author{Umananda Dev Goswami \orcid{0000-0003-0012-7549}}
\email[Email: ]{umananda2@gmail.com}

\affiliation{Department of Physics, Dibrugarh University,
Dibrugarh 786004, Assam, India}

\begin{abstract}
One of the most favourable extensions of General Relativity is the $f(R)$ 
gravity. $f(R)$ gravity in Palatini formalism can be a realistic alternative 
to the dark energy problem. In this work we study a recently introduced dark 
energy $f(R)$ gravity model along with two other models in cosmological 
perspectives under the Palatini formalism. First, we study the cosmic expansion
history of these models with the help of the important cosmographic 
parameters, such as the Hubble parameter, luminosity distance, effective 
equation of state etc. This study shows that the new model behaves similarly 
with the other two models as well as with the $\Lambda$CDM model in some 
respects in the early or very early phases of the universe. It could predict 
the present accelerated expansion of the universe somewhat differently from 
the other models with a peculiar future history of the universe. Within a 
constrained range of parameters all models show a good agreement with the 
Union2.1 luminosity distance data. However, the new model shows a quite 
satisfactory agreement in the whole range of its allowed parameters than 
that of the other two models. We also obtain cosmological constraints on 
these models from the Observed Hubble Data. Further, models have been tested 
by using $Om(z)$ test and statefinder diagnostics. These diagnostics 
especially, the statefinder diagnostic shows that the evolutionary differences 
between these models are distinct. The evolutionary trajectories of the new 
model are completely different from the other two models we have considered.
\end{abstract}

\maketitle

\section{Introduction}

Einstein's General Relativity (GR) is the pioneer of modern cosmology. GR is 
an excellent theory to study the expansion history of the universe. It predicts
so many things and they are in excellent agreement with the observations. 
However, the observations of type \uppercase\expandafter{\romannumeral1}a 
supernovae (SN\uppercase\expandafter{\romannumeral1}a) 
\cite{Perlmutter1999,Riess1998} show that the universe is currently at an 
accelerated expansion phase. This observed result is in contradiction with our 
expectation from the behavior of ordinary matter and it could not be 
explained on the basis of GR. 
One of the best possible explanations of the current accelerated expansion of 
the universe is to assume the existence of an exotic form of energy in the 
universe known as dark energy, which has the characteristic of negative 
pressure. For the simplest explanation of dark energy, initially abandoned 
the well known Einstein's cosmological constant $\Lambda$ has been 
reintroduced as the energy of the vacuum state of the universe 
\cite{Sahni2000,Copeland2006}. In this model the (vacuum) energy density 
$\rho_{\Lambda} = \Lambda/8\pi G$ ($c=1$) is a constant in time and hence the 
universe evolves with an equation of state $w = -1$. But there are two main 
issues with this model: problems of the fine tuning and the cosmic coincidence.
The problem of fine tuning is related to the huge difference between the 
observed value of $\rho_{\Lambda}$ and its calculated value from the concept 
of zero-point energy fluctuations of QFT. This calculated value is found to 
be $\sim 10^{121}$ orders of magnitude larger than the observed value if one 
uses the energy scale as the Planck scale. Even if we use a much lower QCD 
energy scale, this difference would be $\sim 10^{44}$ orders of magnitude 
\cite{Copeland2006}. Thus, this is indeed a severe problem in this field of 
study. On the other hand, the cosmic coincidence problem is based on the 
fact that the dark energy density is approximately equal to the dark matter 
density of the universe today and we do not have a clear answer to the  
reason behind this equality \cite{Velten2014}. These problems motivated the 
researchers to look for some different approaches to solve the dark energy 
issue. Moreover, dark energy is attributed to a huge fraction of the 
universe which is still mysterious to us. This dominant component of the 
universe not only has a large negative pressure, but also it does not 
cluster/behave as like the other ordinary matters/energies of the universe do. 
That is it does not resemble other known matter energy. 

Since dark energy is not directly detected experimentally yet, and hence without a confirmed 
experimental signature, it keeps the window open to other alternative choices 
without such an enigma. However, there are some recent studies showing possibility of detection of dark energy \cite{sunny2, sunny3}. Among alternative possibilities to explain the 
accelerated expansion of the universe is the modification of gravity law in 
such a way that it behaves as standard GR in strong gravitational regimes and 
behaves as repulsive force in the low density cosmological scale. For the 
modification of gravity theory there are two choices: one is to alter the 
matter part and the other is to modify the curvature part. The modification of 
the matter energy part basically involves the introduction of a better dark 
energy model to overcome the drawbacks of the $\Lambda$ model. Analysis done 
on the SN\uppercase\expandafter{\romannumeral1}a data before some twenty years 
ago indicates that a time dependent dark energy can give a better fit than the 
previous cosmological constant model. In view of this the dubbed quintessence 
got significant attention from the researchers, where they used dynamical 
scenarios of dark energy instead of using $w$ as a constant in time 
\cite{Ratra1988,Peebles1988,Ostriker1995}. In another study based on the 
latest observations, it was found that the dynamical dark energy model is 
preferred at a 3.5$\sigma$ significance level \cite{Zhao2017}. Apart from 
these, phantom \cite{Caldwell2003,Singh2003}, quintom \cite{Guo2005,Feng2005}, 
ghost condensates \cite{Arkani-Hamed2004,Piazza2004} etc.\ dark energy models 
have got sufficient attention of the researchers. Further, with an attempt to 
unify the dark energy and dark matter the Chaplygin gas has been introduced 
\cite{Bento2002,Bento2004}. Similarly, there are some other models also 
which are introduced as an explanation of dark energy as well as dark matter 
without modifying the geometry of spacetime \cite{quin01}.

In the other choice, the geometry part of the field equation is modified. The 
$f(R)$ gravity is one of such theories in which modification of spacetime 
curvature takes place by replacing the Ricci curvature scalar in E-H action 
with a general function of it. For $f(R)$ gravity, there are two main 
approaches to obtain the field equations. The first one is the so called 
Metric formalism, which is obtained by the variation of the action with 
respect to the metric. In the second approach, which is known as the Palatini 
formalism, both the connection and the metric are considered as independent 
variables. In recent days the Palatini formalism seems to get more importance 
because of a couple of advantages over the Metric formalism. Here the 
field equations are a set of second order partial differential equations plus 
an equation involving connection, which is trivial to solve; in contrast to the 
4th order differential equations in metric formalism. Moreover, it was brought 
to notice by Dolgov and Kawasaki that the fourth order equations in metric 
formalism suffers from a very serious instability problem \cite{Dolgov2003,Soussa2004}. But in the case of Palatini formalism, the field equations being only 
second order are free from such instabilities \cite{Meng2003,Meng2004}. 
 It 
was also discovered that some $f(R)$ gravity models in this formalism can not 
produce a standard matter-dominated era followed by an accelerated expansion 
\cite{Amendola2007,Amendola2007P}.
 Another important issue with the metric formalism $f(R)$ gravity models is that some 
models do not pass the solar system tests. For example, models of the type 
$f(R) = R-\beta/R^n$, $f(R) = \alpha R^n$ etc.\ do not pass the solar system 
tests \cite{Chiba2003}.  However, there are several $f(R)$ gravity models which 
can easily pass solar system tests and also can describe the matter dominated stage followed by the present accelerating stage of the universe in metric formalism. Some of them are: 
Starobinsky model \cite{Starobinsky}, Hu-Sawicki model \cite{hu01} etc. In addition to them, there are some other promising models: a new 
model defined in \cite{gogoi01}, another model defined in \cite{sunny1} etc.  Apart from this, there is an issue with 
the correct Newtonian limit also \cite{Sotiriou2006,Sotiriou2006G}.
 But in the case of Palatini formalism we do 
not have to check the solar system tests separately as all $f(R)$ gravity 
models pass solar system tests in this formalism. Moreover, in this formalism, 
the correct Newtonian limit is easily recovered \cite{Sotiriou2006G}. It was 
also shown that the selected models in Palatini formalism can produce the 
sequence of radiation-dominated, matter-dominated and late time accelerating 
phases successfully \cite{Fay2007}. Due to these positive points of the 
formalism, we shall proceed with the Palatini formalism $f(R)$ gravity 
hereafter. For a more details on similar studies in modified gravity readers may follow the Ref.s \cite{n1, n2, n3, n4}.

So far a few models are tested in Palatini formalism to study the expansion 
history of the universe. Here in this work we try to explain the nature of 
past, present and future acceleration/deceleration of the universe with the 
help a new $f(R)$ gravity model \cite{gogoi01} in comparison with the power law 
model of $f(R)$ gravity and a model of the form $f(R) =  R + \alpha R^n$. The 
reason behind choosing these models are as follows. The power law model is a 
very simple extension of GR. Although this model does not behave well in the 
solar system regime in metric formalism, in Palatini formalism it is free from 
such issues. In a very recent study, the polarization modes of gravitational 
waves were studied in this model and for a special case of the model, it 
behaves very uniquely in terms of gravitational wave polarization modes than 
from other $f(R)$ gravity models \cite{gogoi02}. The model of the form 
$f(R) =  R + \alpha R^n$ contains the Starobinsky's inflation model
\cite{Starobinsky_inflationary} as a subset. Apart from this, some previous studies on such types of models show 
significant results \cite{hough, Cao2018}. Finally, the new $f(R)$ 
gravity model defined in Ref.\ \cite{gogoi01} is not studied in cosmological 
perspectives yet. For the rest of this paper, we shall use the term 
``Gogoi-Goswami model" to denote this new model specifically. 
 It should be pointed out that in this study the first two models are 
considered as the standard references for the comparative analysis of the new
model results as these two models are studied many times in the similar 
context. For instance, in Ref.\ \cite{Alle04}, the power law model of $f(R)$ 
gravity has been used and it is shown that such models are able to explain the 
current acceleration of the universe with a suitable choice of the model 
parameters. Authors have also introduced a new Lagrangian which is a function 
of the inverse of $\sinh(R)$ and can effectively show the current acceleration 
of the universe. In another study, the power law model was used in the 
Friedmann-Robertson-Walker metric and it was found that first-order nonlinear 
gravity can survive SNIa and baryon oscillation tests \cite{Boro06}. In a 
recent study, polynomial $f(R)$ gravity is considered in dynamical system 
approach, which showed that the Starobinsky $f(R)$ gravity in the Jordan frame 
and Einstein frame are not physically equivalent \cite{Szy18}. Being motivated 
from these results, in this work, we have considered our new dark energy 
$f(R)$ gravity model, as mentioned above which has not been studied earlier in 
cosmological perspectives. Therefore, we believe, a study of this model in this direction will provide some useful insights to its behaviour and viability in 
the respective areas. Apart from this, the study will also provide a constraint on this new model from the available observational data. We shall also study 
the behaviour of the models in the $Om(z)$ test and the statefinder diagnostics, and then will compare the results to see the viabilities and uniqueness of 
this new model.

The rest of the paper is organized as follows. In section \ref{sec02}, we 
briefly explain the derivation of field equations in Palatini formalism and 
find out the connection for the $f(R)$ gravity models in this formalism. In 
section \ref{sec03}, we define the basic mathematical formulations required 
for the study along with the necessary physical and cosmographical parameters. 
In section \ref{sec04}, we define the $f(R)$ gravity models used in this study 
and then calculate the cosmographical parameters for these models. Also in this 
section the numerically calculated cosmographical parameters are compared with 
the available data and corresponding results are discussed. We constrain these 
models using the observed Hubble data in section \ref{sec05}. We used the 
$Om(z)$ test and statefinder diagnostics in section \ref{sec06} to study the 
cosmological behaviours of these models. Finally we conclude the paper with a 
concise discussion of the results in section \ref{sec07}.


\section {Palatini formalism of $f(R)$ gravity} \label{sec02}
The simplest class of modified gravity models is the $f(R)$ gravity models 
in which Einstein gravity is modified by replacing the Ricci curvature scalar 
$R$ by an arbitrary curvature function $f(R)$. The action for the $f(R)$ 
gravity models in the Jordan frame is given by \cite{Sotiriou,Nojiri,Olmo}
\begin{equation}
S=\frac{1}{2\kappa^2}\int d^4x\sqrt{-g}f(R)+S_m (g_{\mu\nu}, \Psi),
\label{eq1}
\end{equation}
where 
$\kappa^2=8\pi G$, $S_m$ is the standard action for matter fields and $\Psi$
represents the matter fields collectively. Throughout this paper we shall work
only in the Jordan frame. The approach to the Einstein frame action can be
found in \cite{Goswami}.

In the Palatini approach, the metric $g_{\mu\nu}$ and the torsion-free 
connection $\Gamma_{\mu\nu}^{\sigma} $ are treated as independent variables.
However, the matter action $S_m$ is assumed to not depend on the independent
connection, but depend only on the metric and the matter fields. This is the
main feature of the Palatini formalism as this assumption is required to derive
Einstein's equations from the action (\ref{eq1}) \cite{Sotiriou}. Moreover,
since the connection is dynamical, we do not need to assume any priori symmetry 
in its lower indices \cite{Olmo}.  Thus, while we vary the action to obtain the 
field equations we consider $\Gamma_{\mu\nu}^{\alpha} \neq 
\Gamma_{\nu\mu}^{\alpha}$. However, for more complicated or higher order Ricci type Lagrangians, one needs to assume a priori that the connection variable is symmetric if a reasonable field equation for the connection is expected \cite{Boro98}. Now varying the action 
(\ref{eq1}) with respect to the metric $g_{\mu\nu}$ we get \cite{Sotiriou,Olmo},
\begin{equation}
R_{(\mu\nu)}f_R - \frac{1}{2}g_{\mu\nu}f(R) = \kappa^2\, T_{\mu\nu},
\label{eq2}
\end{equation}
where $f_R$ denotes the derivatives of $f(R)$ with respect to $R$, $(\mu\nu)$
represents symmetrization over the indices $\mu$ and $\nu$, and $ T_{\mu\nu}$ 
is the energy-momentum tensor as given by
\begin{equation}
T_{\mu\nu}=\frac{-2}{\sqrt{-g}}\frac{\delta S_m}{\delta g_{\mu\nu}}.
\label{eq3}
\end{equation}
It is to be noted that when $f(R) = R$ the Eq.\ (\ref{eq2}) yields Einstein's
equations, i.e. in the limit of $f(R) = R$ the Palatini formalism leads to GR.
      
Again varying the action with respect to $\Gamma_{\mu\nu}^{\sigma}$ we get,
\begin{equation}
\nabla_{\alpha}\left[f_R \sqrt{-g}g^{\mu\nu}\right]-\frac{1}{2}\nabla_{\sigma}\left[f_R\sqrt{-g}g^{\sigma\mu}\right]{\delta_\alpha}^\nu-\frac{1}{2}\nabla_{\sigma}\left[f_R\sqrt{-g}g^{\sigma\nu}\right]{\delta_\alpha}^{\mu}=0,
\label{eq4}
\end{equation}
where $\nabla_{\mu}$ represents the covariant derivative. By taking the trace 
it is easy to show that the Eq.\ (\ref{eq4}) is equivalent to  
\begin{equation}
\nabla_\alpha\left[f_R\sqrt{-g}g^{\mu\nu}\right]=0.
\label{eq5}
\end{equation}
In the GR limit this equation becomes the definition of the Levi-Civita 
connection, i.e.\ in the Palatini formalism for GR, the independent 
connection turns out to be the Levi-Civita one. This is due to the dynamical 
feature of the connection $\Gamma_{\mu\nu}^\alpha$ as mentioned above. To find
the solution of Eq.\ (\ref{eq5}), we define a new metric
\begin{equation}
h_{\mu\nu} = g_{\mu\nu}f_R.
\label{eq5a}
\end{equation}
In terms of this conformal metric to $g_{\mu\nu}$, the Eq.\ (\ref{eq5}) 
transformed into
\begin{equation}
\nabla_\alpha\left[\sqrt{-h}h^{\mu\nu}\right]=0.
\label{eq5b}
\end{equation}
This equation gives the definition of the Levi-Civita connection of $h_{\mu\nu}$
and the algebraic solution of the equation leads to the connection of the 
from: 
\begin{equation}
\Gamma_{\mu\nu}^{\alpha} = \frac{h^{\alpha\sigma}}{2}\left(\partial_\mu h_{\nu\sigma} + \partial_\nu h_{\mu\sigma} - \partial_\sigma h_{\mu\nu}\right).
\label{eq5c}
\end{equation} 
In terms of $g_{\mu\nu}$ it can be expressed as
\begin{equation}
\Gamma_{\mu\nu}^{\alpha}=\left\{\!\!\!\begin{array}{c}\begin{array}{c}\alpha \end{array}\\\begin{array}{cc}\mu&\nu\end{array} \end{array}\!\!\!\right\}+\frac{1}{2f_R}\left(\delta_{\nu}^{\alpha}\partial_{\mu}f_R +\delta_{\mu}^{\alpha}\partial_{\nu}f_R -g^{\alpha\sigma}g_{\mu\nu}\partial_{\sigma}f_R\right),
\label{eq6}
\end{equation}
where $\left\{\!\!\!\begin{array}{c}\begin{array}{c}\alpha \end{array}\\
\begin{array}{cc}\mu&\nu\end{array} \end{array}\!\!\!\right\}$ is the 
Christoffel symbol of 2nd kind of the metric $g_{\mu\nu}$. It is clear from
this equation that the connection, and hence the gravitational fields are
described by the metric and also by the proposed $f(R)$ function.   

\section{Cosmological equations} \label{sec03}
In the large scale the universe is looking homogeneous at every point and 
isotropic in all directions. So, as a simplest model we may consider the
homogeneous and isotropic flat universe described by the 
Friedmann-Lema\'itre-Robertson-Walker (FLRW) metric:
\begin{equation}
ds^2 = -\,dt^2 + a^2(t)\,d\mathbf{r}^2,
\label{eq6a}
\end{equation}
where we used the metric convention as $(-,+,+,+)$ and $a(t)$ is the 
cosmological scale factor. Similarly for an ideal situation we consider that
the universe is filled with a perfect fluid which is the source of curvature.
If $\rho$ is the energy density and $p$ is the pressure of such fluid, then its 
energy-momentum tensor can be written as
\begin{equation}
T^\mu_\nu = diag(-\rho, p, p, p).
\label{eq6b}
\end{equation}
For our universe define by Eqs.\ (\ref{eq6a}) and (\ref{eq6b}), we may obtain 
the generalized Friedmann equation from the Eqs.\ (\ref{eq2}) and (\ref{eq6}), 
which can expressed in terms of redshift $z$ as \cite{Santos} 
\begin{equation}
\frac{H^2}{H_{0}^{2}}=\frac{3\,\Omega_{m0}\left(1+z\right)^3+6\,\Omega_{r0}\left(1+z\right)^4+\frac{f\left(R\right)}{H_{0}^{2}}}{6f_R\zeta^2},
\label{eq7}
\end{equation}
where
\begin{equation}
\zeta=1+\frac{9f_{RR}}{2f_R}\frac{H_{0}^{2}\,\Omega_{m0}\left(1+z\right)^3}{Rf_{RR}-f_{R}}.
\label{eq8}
\end{equation}
Here $\Omega_{m0}$ and $\Omega_{r0}$ are the present day value of the matter 
and radiation density parameters respectively. The trace of the field Eq.\ 
(\ref{eq2}) gives: 
\begin{equation}
Rf_R - 2f(R) = \kappa^2 T.
\label{eq8a}
\end{equation}
Using the energy-momentum tensor given in Eq.\ (\ref{eq6b}), this equation
can be written as 
\begin{equation}
Rf_R - 2f(R) =  - \kappa^2\rho (1-3w),
\label{eq8b}
\end{equation}
where $w = p/\rho$ is the cosmological equation of state. For the radiation 
dominated universe $w = 1/3$ and hence the above equation reduces to such a
universe as
\begin{equation}
Rf_R - 2f(R) = 0.
\label{eq8c}
\end{equation} 
This equation can be directly obtained from the Eq.\ (\ref{eq8a}) as 
$T= 0$ for the radiation dominated universe. However, we will not consider 
this equation further as it creates an inconsistent situation for the field
equations for most of the $f(R)$ models \cite{Sotiriou}. As the radiation
dominated era of the universe has insignificant contribution to the present
stage of the universe it will not affect our result. On the other hand for
the matter dominated universe $w = 0$ and if we take 
$\rho_m = \rho_{m0}\, (1+z)^3$ and then substituting 
$\kappa^2\rho_{m0} = 3H_0^2\,\Omega_{m0}$, we may write the Eq.\ (\ref{eq8b}) 
for the matter dominated universe in the form:
\begin{equation}
Rf_R - 2f(R) = -\,3\,H_0^2\,\Omega_{m0} (1+z)^3.
\label{eq9}
\end{equation}   
This equation will provide us a condition to restrict the parameter of a 
given $f(R)$ model.

Apart from the Hubble parameter $H(z)$, which is given by the Eq.\ (\ref{eq7}), 
other two important cosmographic parameters to study the expansion history of
the universe are effective equation of state $\omega_{eff}(z)$ and deceleration 
parameter $q(z)$. These parameters can be studied from the following 
equations along with the Eq.\ (\ref{eq7}) as given by \cite{Sotiriou,Santos}
\begin{equation}
\omega_{eff}(z) =-1+\frac{2\left(1+z\right)}{3H(z)}H^{\prime}\left(z\right),
\label{eq9a}
\end{equation}

\begin{equation}
q\left(z\right)=\frac{\left(1+z\right)}{H\left(z\right)}H^{\prime}\left(z\right)-1,
\label{eq9b}
\end{equation}
where a prime denotes differentiation with respect to $z$. Another important 
parameter is the distance modulus $D_m(z).$ It is possible to express the 
distance modulus in terms of an $f(R)$ gravity model in the following way:
\begin{equation}\label{Dm}
D_m = 5 \log_{10} \left(\frac{\sqrt[3]{H_0^4\, \Omega _{m0}^2 \left(2 f\left(R_1\right)-R_1 f'\left(R_1\right)\right)} }{\sqrt[3]{3}\, H_0^2\, \Omega _{m0}} \int_{R_0}^{R_1} \frac{69288.2 \; H_0^2\, \Omega _{m0} \left(f'(R)-R f''(R)\right)}{H(R) \left(H_0^4\, \Omega _{m0}^2 \left(2 f(R)-R f'(R)\right)\right)^{2/3}} \, dR\right),
\end{equation}
where $R_1$ is some arbitrary curvature in the past for some fixed value of 
$z$. $R_0$ is the present value of curvature at $z=0.$ In the above 
expression, $H$ is a function of $R$ only. For the rest part of the study, 
we will use $\Omega_{m0} = 0.315,$ and $\Omega_{r0} = 0.000053$ if not 
specified. We consider $H_0 = 68.02$ km s$^{-1}$ Mpc$^{-1}$ \cite{Zhang2021}, 
which is very close to recent Planck's result \cite{Planck2018}.
We shall use this expression to compare the models with the Union2.1 
data \cite{Suzuki2012}.

Since, the forms of different models might increase the complexity of 
obtaining an analytical solution of $R$ in terms of $z$, we may not obtain 
some parameters like $H$ as a function of $z$ directly. To solve this issue, 
we shall express $H$, $\omega_{eff}$ etc.\ as a function of $R$. To do so, 
we shall use the following expressions. $H'(z)$ relates to $dH/dR$ as follows:
\begin{equation}
  H'(z)=\frac{dH}{dR}\frac{dR}{dz}=-\frac{dH}{dR}\frac{9\,\Omega_{m0}H_0^2(1+z)^2}{f''R-f'}.
\end{equation}
$H''(z)$ is expressed as
\begin{equation}
  H''(z)=\frac{dH}{dR}\frac{d^2R}{dz^2}+\frac{d^2H}{dR^2}\left(\frac{dR}{dz}\right)^2,
\end{equation}
where the second derivative of $R$ relating to $z$ is
\begin{equation}
  \frac{d^2R}{dz^2}=\frac{dR}{dz}\left(\frac{2}{1+z}-\frac{f'''R}{f''R-f'}\frac{dR}{dz}\right).
\end{equation}

\section{ $f(R)$ gravity models and properties of cosmic evolution} 
\label{sec04}
Using the field equations and cosmological equations as discuss in the previous
sections, in this section we will study the evolution of universe and 
consequent cosmological implications of the power law model, model of the type 
$R+\alpha R^n$ and the Gogoi-Goswami model as follows:

\subsection{Power law model}
The general power law $f(R)$ gravity model is given by 
\begin{equation}
f(R) = \lambda R^n,
\label{eq10}
\end{equation}  
where $\lambda$ and $n$ are model parameters. Using the Eq.\ (\ref{eq9}) 
on this model for $z=0$, we can find out the restrictions on the model 
parameters as given by 

\begin{equation}
\lambda=-\frac{3\,H_{0}^{2}\,\Omega_{m0}}{\left(n-2\right){R_0}^n}.
\label{eq11}
\end{equation}
It is clear from this expression that the model parameter $n$ should have 
values other than $2$ to avoid the singular value of the parameter $\lambda$. 
In fact it is very easy to show that $n = 2$ is the trivial solution of the
Eq.\ (\ref{eq8c}) for this model. That is for the value of $n = 2$, the 
power-law model specifies the radiation dominated phase of the universe. As the
parameter $\lambda$ depends upon the value of $R_0$ also, so to find the 
expression for $R_{0}$ we proceed as follows:

For the model (\ref{eq10}) the Eq.\ (\ref{eq8}) gives:
\begin{equation}
\zeta = 1 - \frac{3}{2}\frac{n-1}{n}\left(\frac{R_0}{R}\right)^n(1+z)^3.
\label{eq11a}
\end{equation}
For $z=0$, the value of $\zeta$ depends only on the parameter $n$ as given by
\begin{equation}
\zeta_0 = 1 - \frac{3}{2}\frac{n-1}{n}.
\label{eq11b}
\end{equation}     
It is seen that $\zeta$ becomes unity and hence it is $f(R)$ model independent 
when the power-law model takes the value $n =1$. Thus this sets another 
restriction on the parameter $n$ of the model. In fact, this value of $n$ 
corresponds to GR. Finally setting $z=0$ in the Eq.\ (\ref{eq7}) and using the 
Eq.\ (\ref{eq11b}) we can find out the expression for $R_{0}$ for the power law model as
\begin{equation}
R_0=-\frac{3\left(3-n\right)^2 H_{0}^{2}\,\Omega_{m0}}{2n\left[\left(n-3\right)\Omega_{m0}+2\left(n-2\right)\Omega_{r0}\right]}.
\label{eq12}
\end{equation}
This equation also sets another restriction on the parameter $n$ that its value
should be other than $3$, otherwise $R_0$ will vanish.
  
\begin{figure}[h!]
\centerline{
   \includegraphics[scale = 0.3]{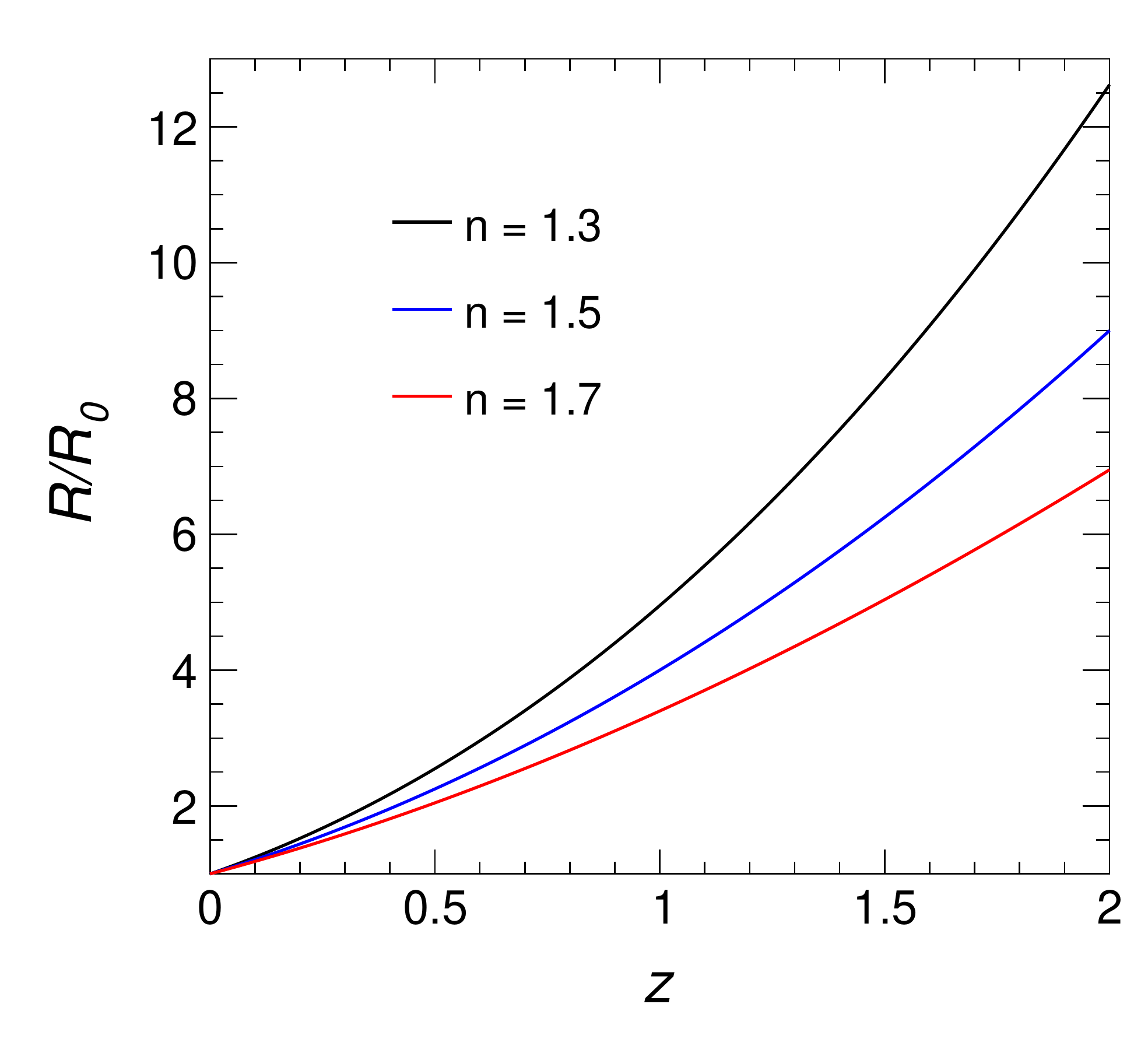}
}
\vspace{-0.2cm}
\caption{Behaviour of scalar curvature $R$ with respect to redshift $z$ for 
three different values of the parameter $n$ of the $f(R)$ gravity power law 
model.}
\label{fig01}
\end{figure}

To find the expression of the scalar curvature function $R$ in terms of 
redshift $z$ we solve the Eq.\ (\ref{eq9}) using the expression (\ref{eq11}) 
for the parameter $\lambda$, which leads to a compact expression of $R$ in the 
form:
\begin{equation}
R(z) =R_0 \left(1+z\right)^{\frac{3}{n}}.
\label{eq13}
\end{equation}
In view of this equation, from Eqs.\ (\ref{eq11a}) and (\ref{eq11b}) it is 
clear that $\zeta = \zeta_0$. So the parameter $\zeta$ is independent of 
cosmological evolution for a given value of the parameter $n$ of the power-law 
model. Fig.\ \ref{fig01} shows the variation of $R$ with respect to $z$ for 
three values of $n$ as given by the Eq.\ (\ref{eq13}). It is seen that the 
slope of the scalar curvature $R$ increases in this model when $n$ approaches 
the GR limit. Also the slope changes significantly for the small variation of 
the value of $n$.

\begin{figure}[h!]
\centerline{
   \includegraphics[scale = 0.3]{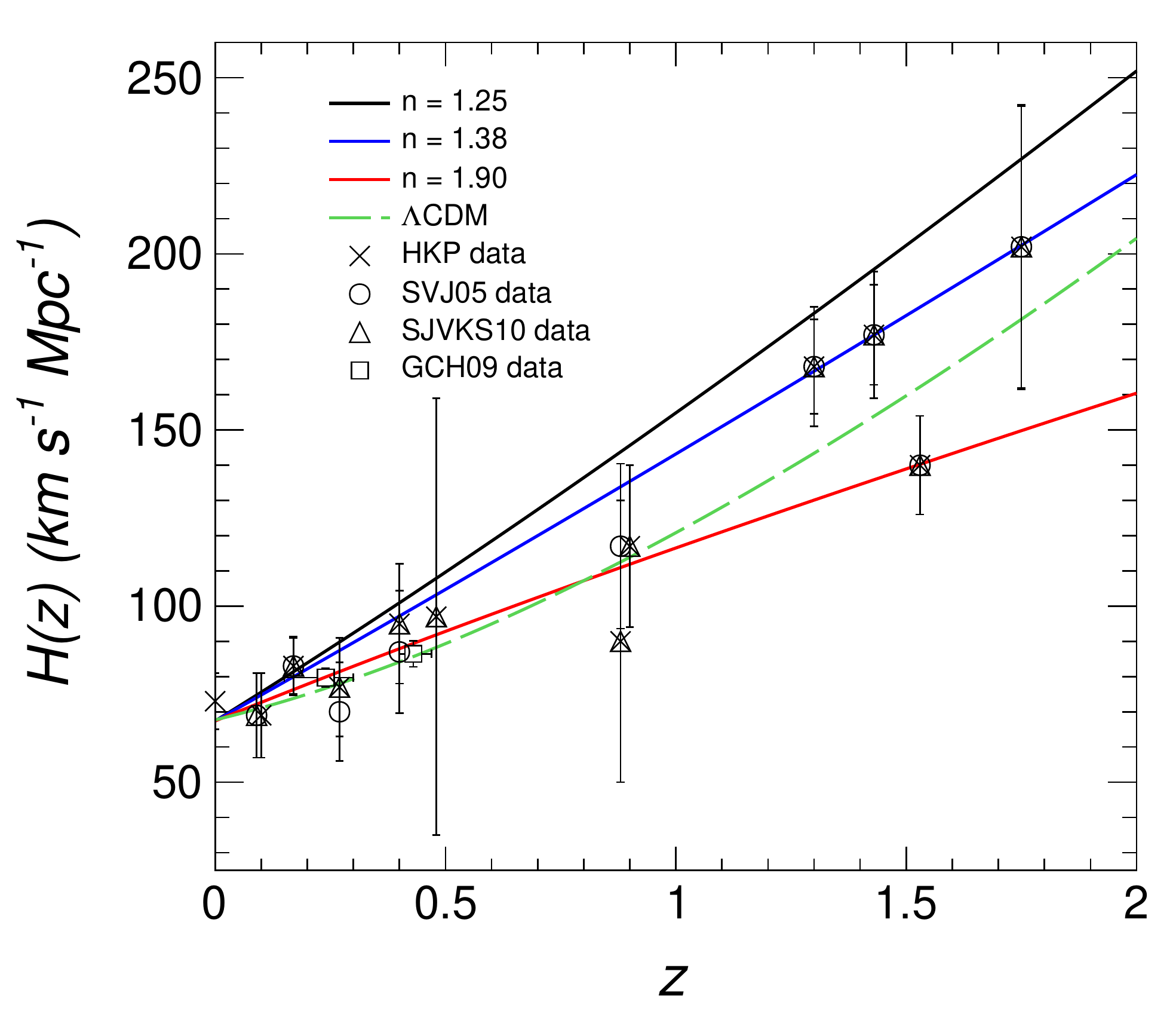}\hspace{0.5cm}
   \includegraphics[scale = 0.3]{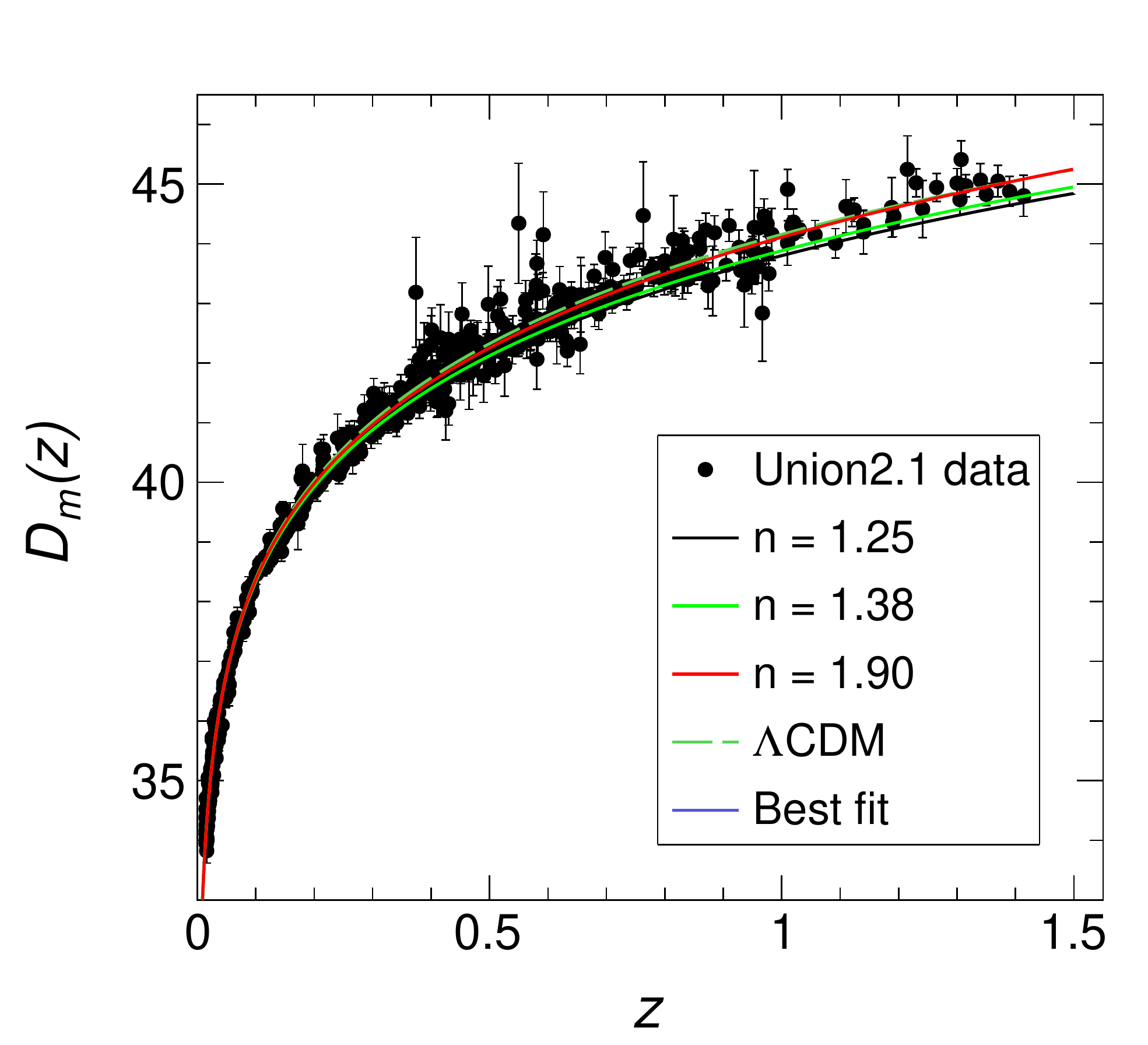}}
\vspace{-0.2cm}
\caption{Behaviours of Hubble parameter $H(z)$ versus redshift $z$ (left panel)
and distance modulus $D_m(z)$ versus $z$ (right panel) for three different 
values of the parameter $n$ of $f(R)$ gravity power law model. The three 
values of $n$ are first obtained by fitting with four sets of observed data 
in the $H(z)$ plot. These values of $n$ are then used to fit the Union$2.1$ 
data \cite{Suzuki2012} in the $D_m(z)$ plot. In both plots the corresponding 
$\Lambda CDM$ model predictions are also shown.}
\label{fig02}
\end{figure}

From the generalized Friedmann Eq.\ (\ref{eq7}) the Hubble parameter $H(z)$ can 
be expressed for the power law model (\ref{eq10}) as  
\begin{equation}
H(z) =\left[-\frac{2nR_0}{3\left(3-n\right)^2\Omega_{m0}}\left\{\left(n-3\right) \Omega_{m0}\left(1+z\right)^{\frac{3}{n}}+2\left(n-2\right)\Omega_{r0}\left(1+z\right)^{\frac{\left(n+3\right)}{n}}\right\}\right]^{\frac{1}{2}}.
\label{eq14}
\end{equation}
As already mentioned, it is also seen from this equation that the value of the 
model parameter $n$ must be different from $3$ to avoid the singular value of
the Hubble parameter to be predicted by this model. Varying this equation for 
$H(z)$ with respect to $z$ we get,
\begin{equation}
H'(z) =\frac{-\,nR_0}{3\left(3-n\right)^2\Omega_{m0}H}\left[\frac{3\left(n-3\right)}{n}\,\Omega_{m0}\left(1+z\right)^{\frac{3}{n}-1}+\frac{2\left(n-2\right)\left(n+3\right)}{n}\,\Omega_{r0}\left(1+z\right)^{\frac{3}{n}}\right]
\label{eq15}
\end{equation}
Expressions of $H(z)$ and $H'(z)$ given by Eqs.\ (\ref{eq14}) and (\ref{eq15}) 
respectively for the power law model can be used to study the other 
cosmological parameters, like the effective equation of state given by the 
Eq.\ (\ref{eq9a}) and the deceleration parameter given in the Eq.\ (\ref{eq9b}).
To understand the behaviour of the Hubble parameter $H(z)$ in terms of $z$ for
different values of $n$, we have plotted $H(z)$ with respect to $z$ for three 
different values of $n$ on the left plot of Fig.\ \ref{fig02}. To choose the 
reliable values of $n$ we have used the four sets of available $H(z)$ data, 
viz., HKP data, SVJ05 data \cite{Simon2005}, SJVKS10 data \cite{Stern2010} and 
GCH09 data \cite{Gaztanaga2009}. Although the SVJ05 data set has been replaced 
already by SJVKS10 data, we have used this data set for reference only 
\cite{Ma2011}. Moreover, in this plot $\Lambda CDM$ model prediction is also
shown. The comparison shows that for smaller values of $n$, the 
theoretical curves deviate from the observational data at higher redshift 
regions ($z>0.5$). The values of $n = 1.38$ and $n=1.90$ show good agreement 
with the observational $H(z)$ data. Again $n=1.90$ shows very close behaviour
with the $\Lambda CDM$ model up to the redshift limit $z\leq 1$. However, we 
used the value $n=1.25$ as the lower limit value of $n$ for the upper limit 
$H(z)$ values at higher redshift regions in this model.

We have used Eq.\ \eqref{Dm} to evaluate the distance modulus $D_m(z)$ for the 
model using the above mentioned three constrained values of $n$. The obtained 
results are compared with the Union$2.1$ data \cite{Suzuki2012} and with
the $\Lambda CDM$ model prediction as shown in the right plot of 
Fig.\ \ref{fig02}. It is observed that for $n=1.90$ the model shows a very 
good resemblance with the experimental data as well as with the $\Lambda CDM$ 
model. With decrease in the value of $n$ the model slightly deviates from the 
$\Lambda CDM$ model as expected. It is because the model approaches GR for 
$n \rightarrow 1$ without a cosmological constant. Further, it should be noted 
that for the other two values of $n$ also, i.e. of $n=1.38$ and $1.25$ the 
power law model predicted values of $D_m(z)$ lay well within the experimental 
data. 

\begin{figure}[h!]
\centerline{
   \includegraphics[scale = 0.3]{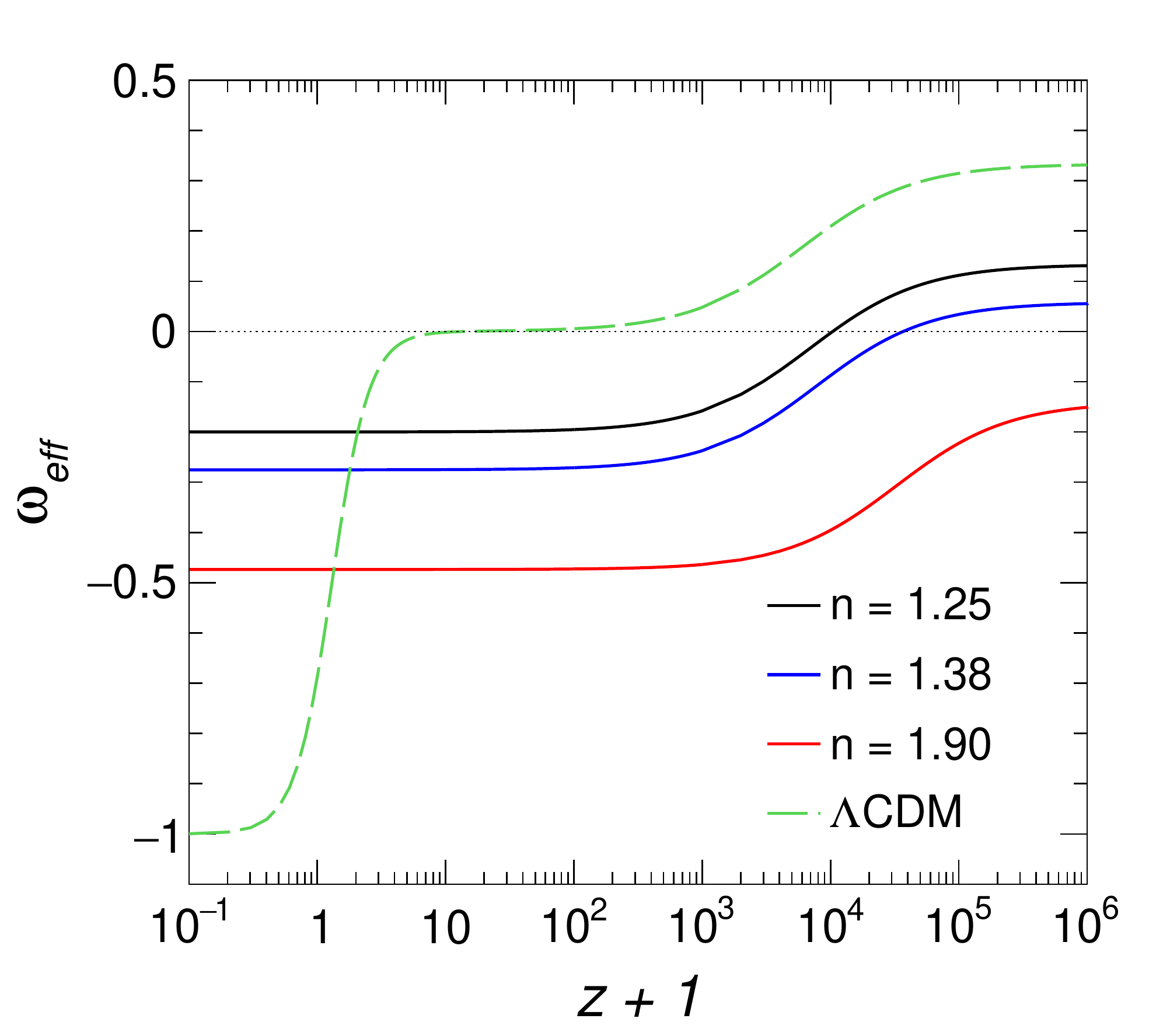}\hspace{0.5cm}
   \includegraphics[scale = 0.3]{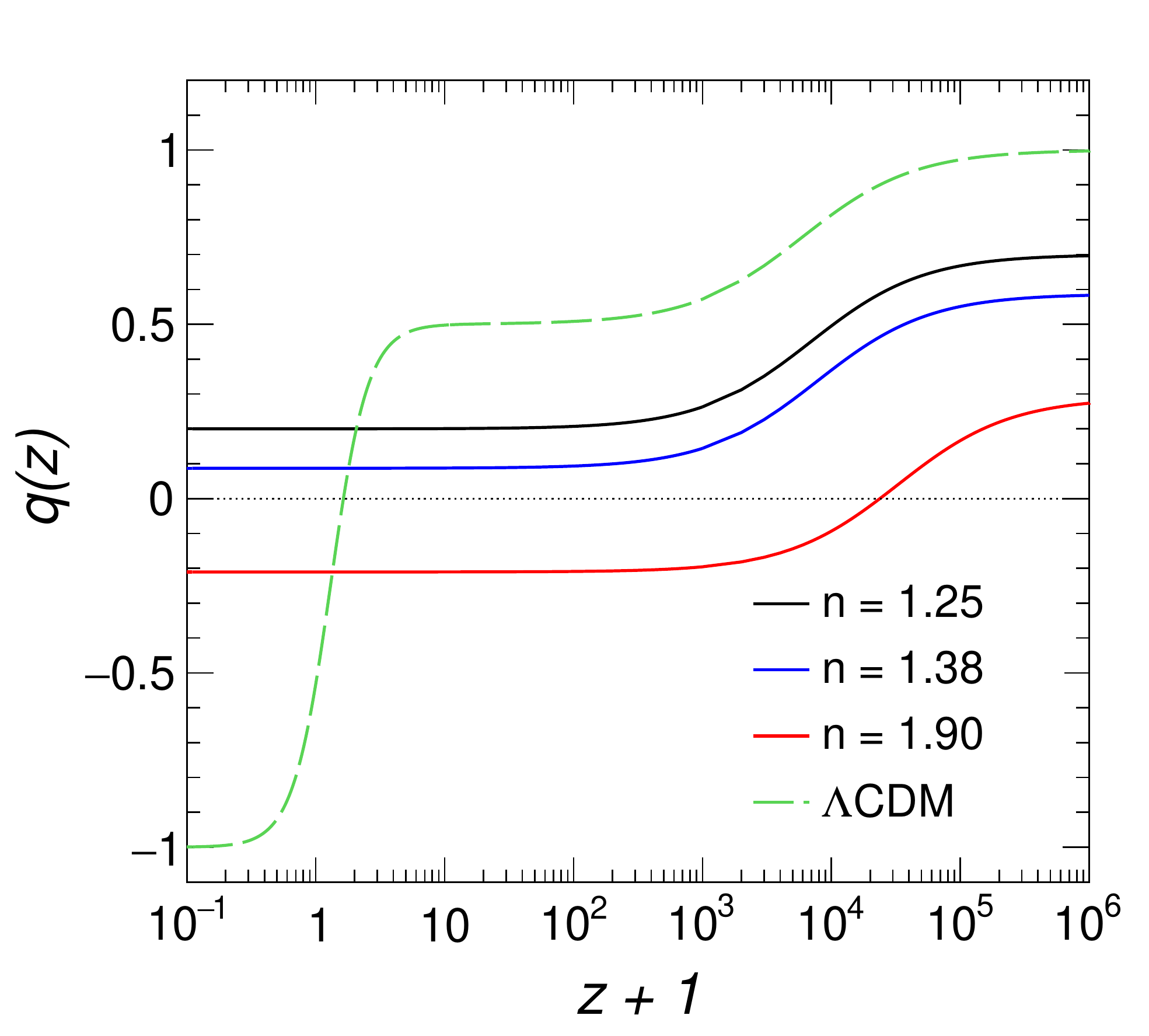}
}
\vspace{-0.2cm}
\caption{Variations of $\omega_{eff}(z)$ versus $z$ (left plot) and 
deceleration parameter $q(z)$ versus $z$ (right plot) for three constrained 
values of the parameter $n$ of $f(R)$ gravity power law model. The predictions 
of these two cosmological parameters by the $\Lambda CDM$ model are also shown 
in the corresponding plots.}
\label{fig03}
\end{figure}

The effective equation state $\omega_{eff}(z)$ and the deceleration parameter 
$q(z)$ given by Eqs.\ \eqref{eq9a} and \eqref{eq9b} respectively are calculated 
with respect to $z$ for the  three constrained values of the parameter $n$ of 
$f(R)$ gravity power law model, using the Eqs.\ \eqref{eq14} and \eqref{eq15}. 
The results of these calculations are plotted respectively on the left and 
right panels of Fig.\ \ref{fig03}. Further, the $\Lambda CDM$ model's 
predictions of these two cosmological parameters are also plotted in the 
corresponding plots. From the $\omega_{eff}$ versus $z$ plot it is seen that 
the power law model based $\omega_{eff}$ deviates significantly from 
$\Lambda$CDM in the present scenario. However, it is observed that the curve 
of $\omega_{eff}$ approaches $\Lambda$CDM with an increase in the value of $n$ 
from $1.25$ to $1.90$ for the present epoch. But, for the higher values of 
$z$, the curve with $n=1.90$ hardly mimics $\Lambda$CDM. One can see that, 
for $n=1.90$ the universe does not have a radiation dominated and matter 
dominated phase of evolution. It has mostly the accelerating phase including
the late time one, but which is not as much as expected. We see the similar 
trend from the $q$ versus $z$ plot. Here it is interesting to see that for 
$n = 1.25$ and $1.38$ the power law model does not predict any accelerating 
phase throughout the history of the universe. These observations indicate 
that the power law model can not efficiently mimic the $\Lambda$CDM model in 
the early universe for large values of $n$ and the current universe for small
values of $n$. For a more detailed observation in this direction, we use the 
$Om(z)$ test and the statefinder diagnostics in section \ref{sec06}.

\subsection{ $f(R) = R + \alpha R^n $ model}
\begin{figure}[htb]
\centerline{
   \includegraphics[scale = 0.3]{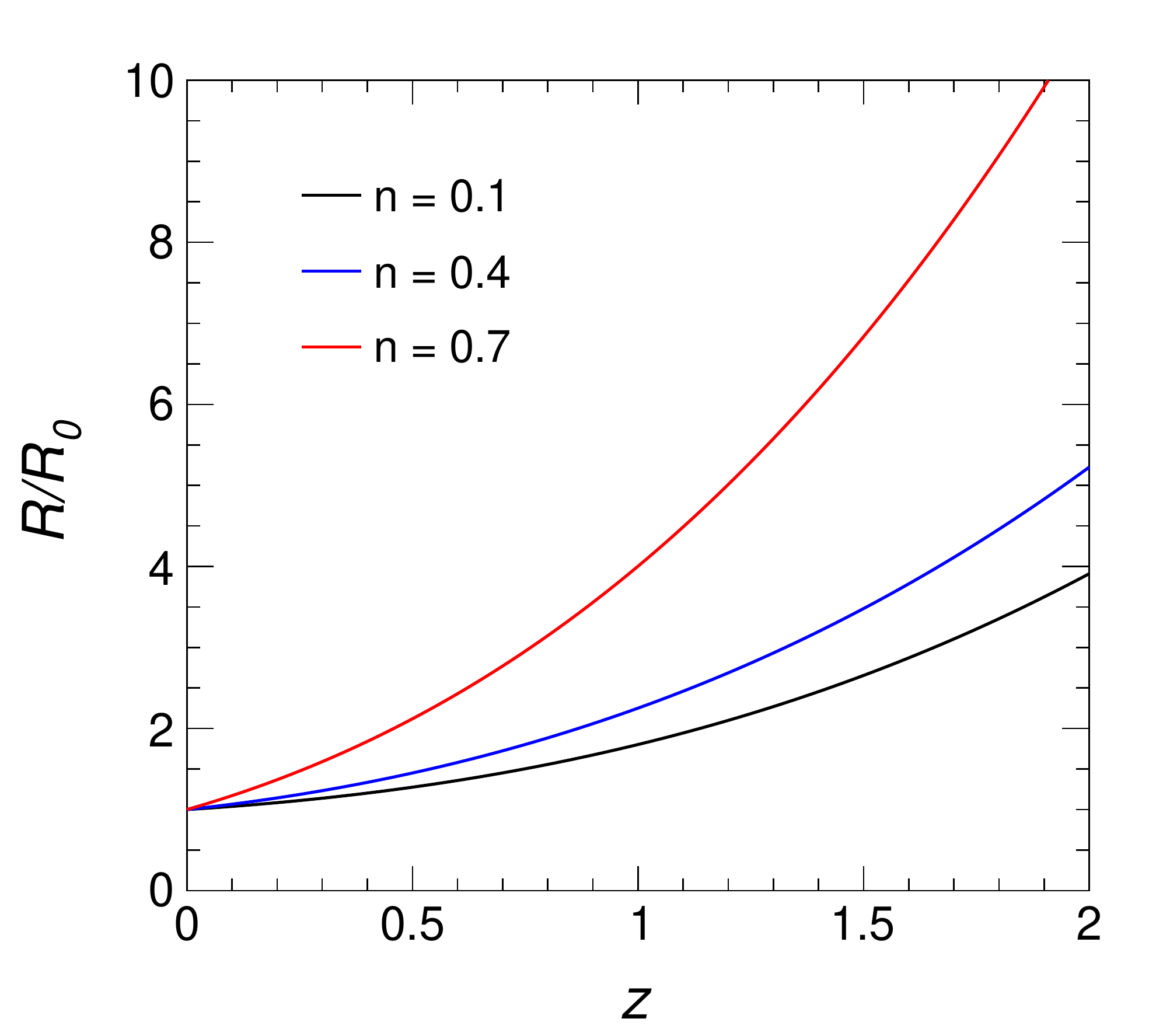}}
\vspace{-0.2cm}
\caption{Variation pattern of $R$ with respect to $z$ for three different 
values of the parameter $n$ of the $f(R) = R + \alpha  H_0^{2 (1-n)} R^n$ 
model.}
\label{fig04}
\end{figure}

This model contains two constant parameters $\alpha$ and $n$. Here the 
parameter $\alpha$ is not dimensionless and its dimension depends on the value
of $n$ used. Accordingly, for the ease of analysis we have to make 
$\alpha$ dimensionless. For this purpose we rewrite the model as
\begin{equation}\label{model2}
f(R) = R + \alpha  H_0^{2 (1-n)} R^n.
\end{equation}
Now using the Eq.~\eqref{eq9} for this model in the limit $z \rightarrow 0$, 
the parameter $\alpha$ can be expressed as
\begin{equation}\label{alpha_model2}
\alpha =\frac{H_0^{2 n-2} R_0^{-n} \left(R_0-3\, H_0^2\, \Omega _{\text{m0}}\right)}{n-2}.
\end{equation}
Similarly, we can use the Eq.\ \eqref{eq7} in the limit $z \rightarrow 0$ to 
get the following expression for this model:
\begin{equation}\label{H0_model2}
\frac{2 R_0 \left(\alpha  H_0^2 n R_0^n+R_0 H_0^{2 n}\right) \left(R_0 H_0^{2 n}-\alpha  H_0^2 (n-2) n R_0^n\right){}^2 \left(H_0^{2 n} \left(3 H_0^2 \left(\Omega _{\text{m0}}+2 \Omega _{\text{r0}}\right)+R_0\right)+\alpha  H_0^2 R_0^n\right)}{3 \left(\alpha  n H_0^{2 n+3} R_0^n \left(9 H_0^2 (n-1) \Omega _{\text{m0}}+2 (n-3) R_0\right)+2 \alpha ^2 H_0^5 (n-2) n^2 R_0^{2 n}-2 R_0^2 H_0^{4 n+1}\right){}^2}=1.
\end{equation}
Using Eq.\ \eqref{alpha_model2} in the above expression, we can solve this
expression for $R_0$ numerically. One can see that the complexity of the above 
equation makes it difficult to have an analytical solution of the present 
background curvature $R_0$. For this model, the relation between $z$ and $R$ 
can be obtained from the Eq.~\eqref{eq9} as given by,
\begin{equation}\label{z_model2}
z=\frac{\sqrt[3]{H_0^4 \Omega _{\text{m0}}^2 \left(R-\alpha  (n-2) H_0^{2-2 n} R^n\right)}}{\sqrt[3]{3} H_0^2 \Omega _{\text{m0}}}-1.
\end{equation}
From this equation we calculate numerically the values of $z$ for a range
of values of $R$ for different values of the model parameter $n$ together with 
the solutions of $R_0$ obtained by using the Eqs.\ \eqref{alpha_model2} and 
\eqref{H0_model2} for the corresponding values of $n$ to see the variation 
pattern of $R$ with respect to $z$ in this model. The results of this 
calculation is shown in Fig.\ \ref{fig04}.  In this case, the scenario is 
opposite to the case of the power law model. Here, with increase in the value 
of the parameter $n$, the slope of the curve increases, predicting deviations 
from the GR. For smaller values of $n$, the curves have smaller slopes and the 
differences are again significant as like in the case of the power law model.

\begin{figure}[h!]
\centerline{
   \includegraphics[scale = 0.3]{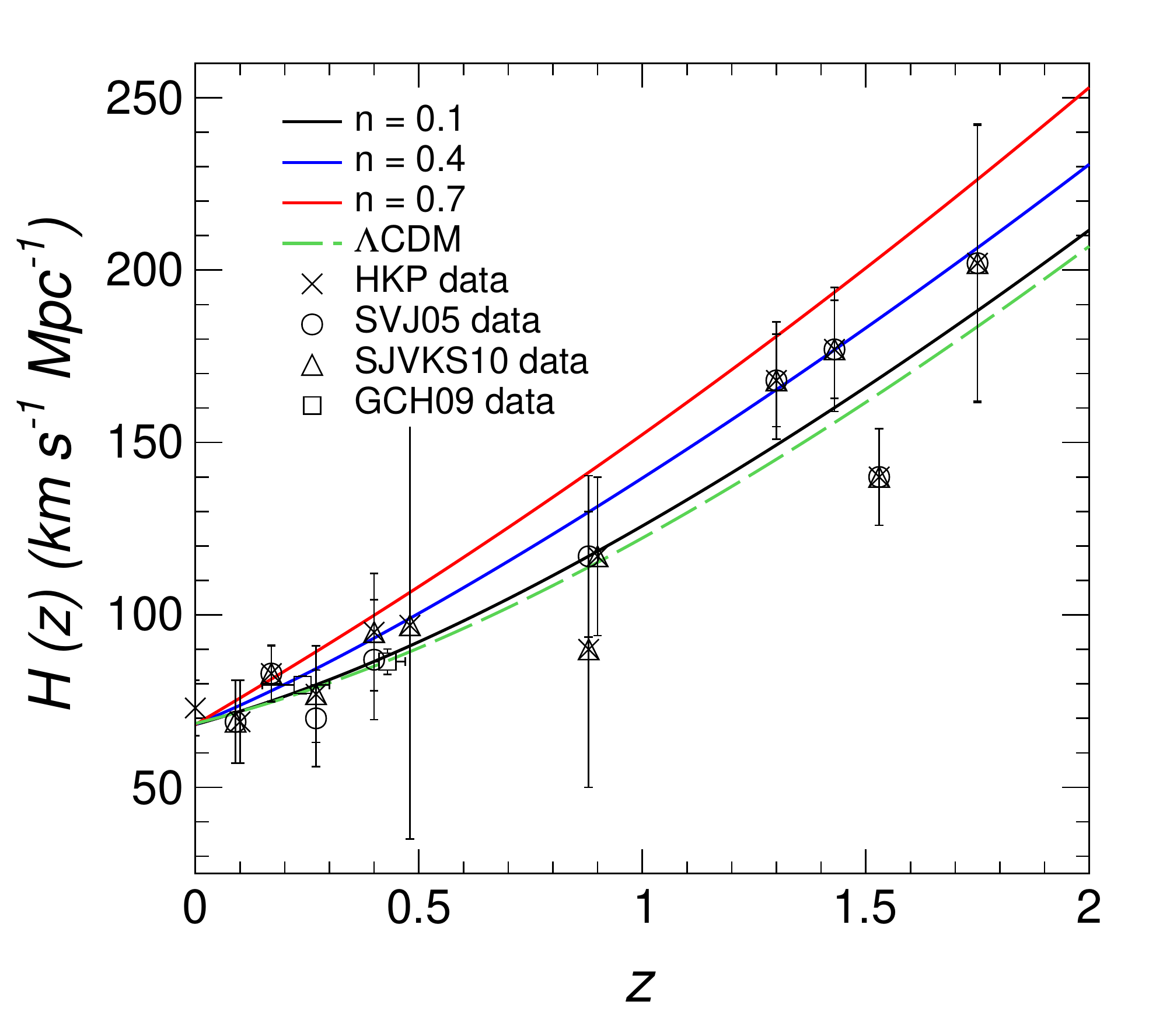} \hspace{0.5cm}
   \includegraphics[scale = 0.3]{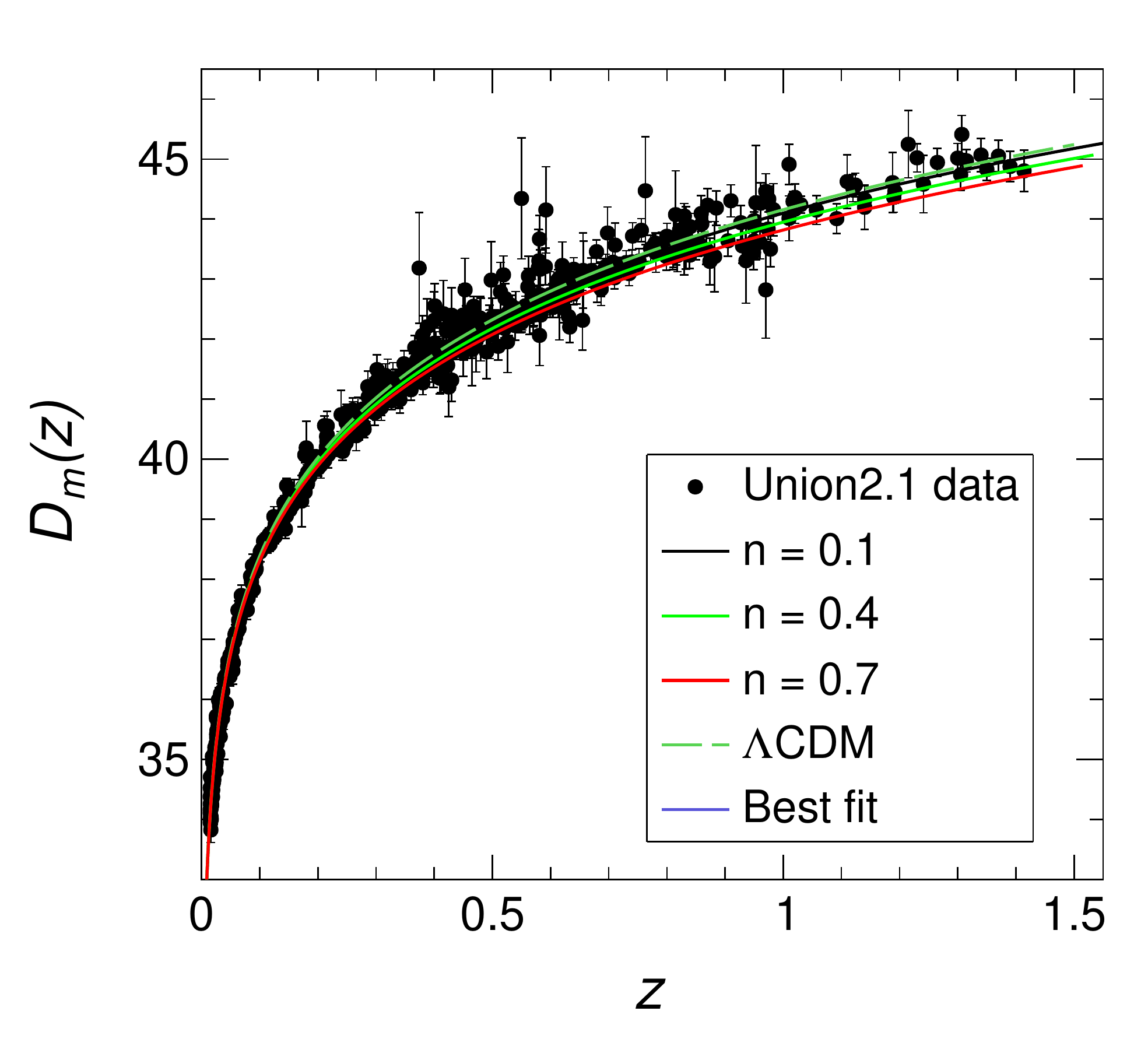}}
\vspace{-0.2cm}
\caption{Variation of Hubble parameter $H(z)$ (left panel) and distance 
modulus $D_m(z)$ (right panel) with respect to $z$ for three different
values of the parameter $n$ of $f(R) = R + \alpha  H_0^{2 (1-n)} R^n$ 
model. Similar to the case of power law model, the three values of $n$ are 
first obtained by fitting with four sets of observed data in the $H(z)$ plot. 
These values of $n$ are then used to fit the Union$2.1$ data \cite{Suzuki2012} 
in the $D_m(z)$ plot. In both plots the corresponding $\Lambda CDM$ model 
predictions are also shown.}
\label{fig05}
\end{figure}

Again, using the Eq.\ \eqref{eq7} we may write the Hubble parameter $H$ for
this model as
\begin{equation}
H = \frac{H_0 \sqrt{R} \sqrt{\alpha  H_0^{-2 n} R^n+\frac{R}{H_0^2}+3 (z+1)^3 \left(\Omega _{\text{m0}}+2 (z+1) \Omega _{\text{r0}}\right)}}{\sqrt{6} \sqrt{\alpha  n H_0^{2-2 n} R^n+R} \left(1-\frac{9 \alpha  (n-1) n (z+1)^3 H_0^{2 n+4} \Omega _{\text{m0}} R^n}{2 \left(\alpha  H_0^2 n R^n+R H_0^{2 n}\right) \left(R H_0^{2 n}-\alpha  H_0^2 (n-2) n R^n\right)}\right)}.
\end{equation}
Now, using the expression for $z$ in this equation it is possible to get $H$ 
as a function of the scalar curvature $R$ from this equation. Finally, all 
these expressions can be used to study the cosmological behaviours of this 
model. With an objective similar to the power law model, we have plotted 
$H(z)$ with respect to $z$ for this model also and compared the model 
predictions with the HKP data, SVJ05 data, SJVKS10 data and GCH09 data as
shown in the left plot Fig.\ \ref{fig05}. This plot shows that the model is in 
good agreement with the observational data for the parameter around $n=0.4$. 
With an increase in the value of $n$ from $0.1$, we observe that the model 
deviates from $\Lambda$CDM model in a significant manner. So, these 
observations predict that the model can be a good alternative of $\Lambda$CDM 
model for smaller $n$ values. Moreover, $n=0.1$ and $0.7$ can be considered 
as lower and upper limits of values of $n$ corresponding to the lower and 
upper limits of values of observed $H(z)$.

Also we use Eq.\ \eqref{Dm} as in case of the power law model to calculate 
numerically the distance modulus for this model using the above three 
constrained values of $n$ and results are shown in the right plot of 
Fig.\ \ref{fig05}. We compare these results with the Union2.1 data and we see 
that the model is in good agreement with the observational Union2.1 data for
all three values of $n$, i.e.\ $n=0.1, 0.4$ and $0.7$. However, it is clear 
that the model shows a better outcome for $n=0.1$. With the increase in the 
value of $n$ from $0.1$, we see a tendency of deviation of the model 
prediction from the observational Union2.1 data.

\begin{figure}[h!]
\centerline{
   \includegraphics[scale = 0.3]{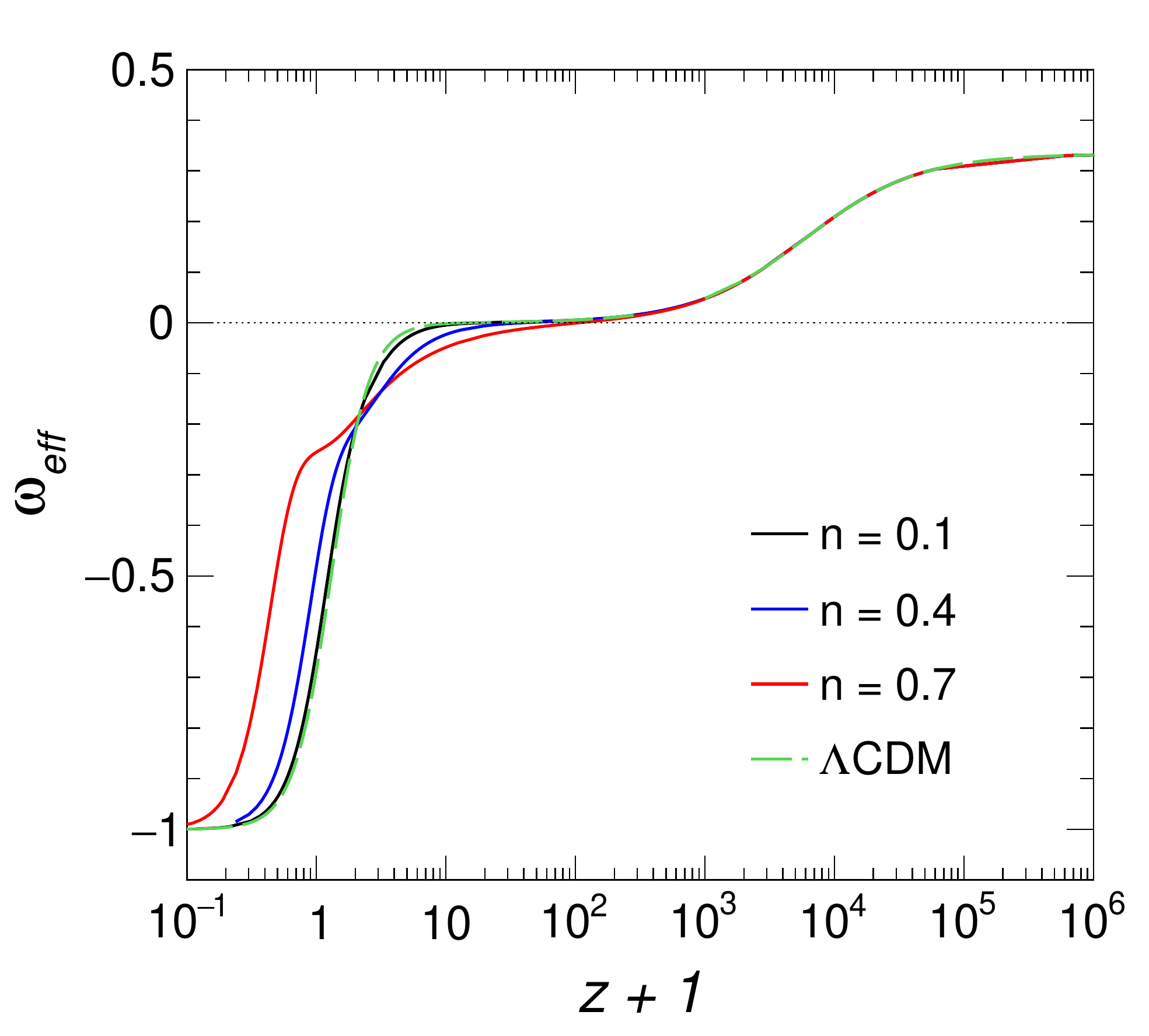}\hspace{0.5cm}
   \includegraphics[scale = 0.3]{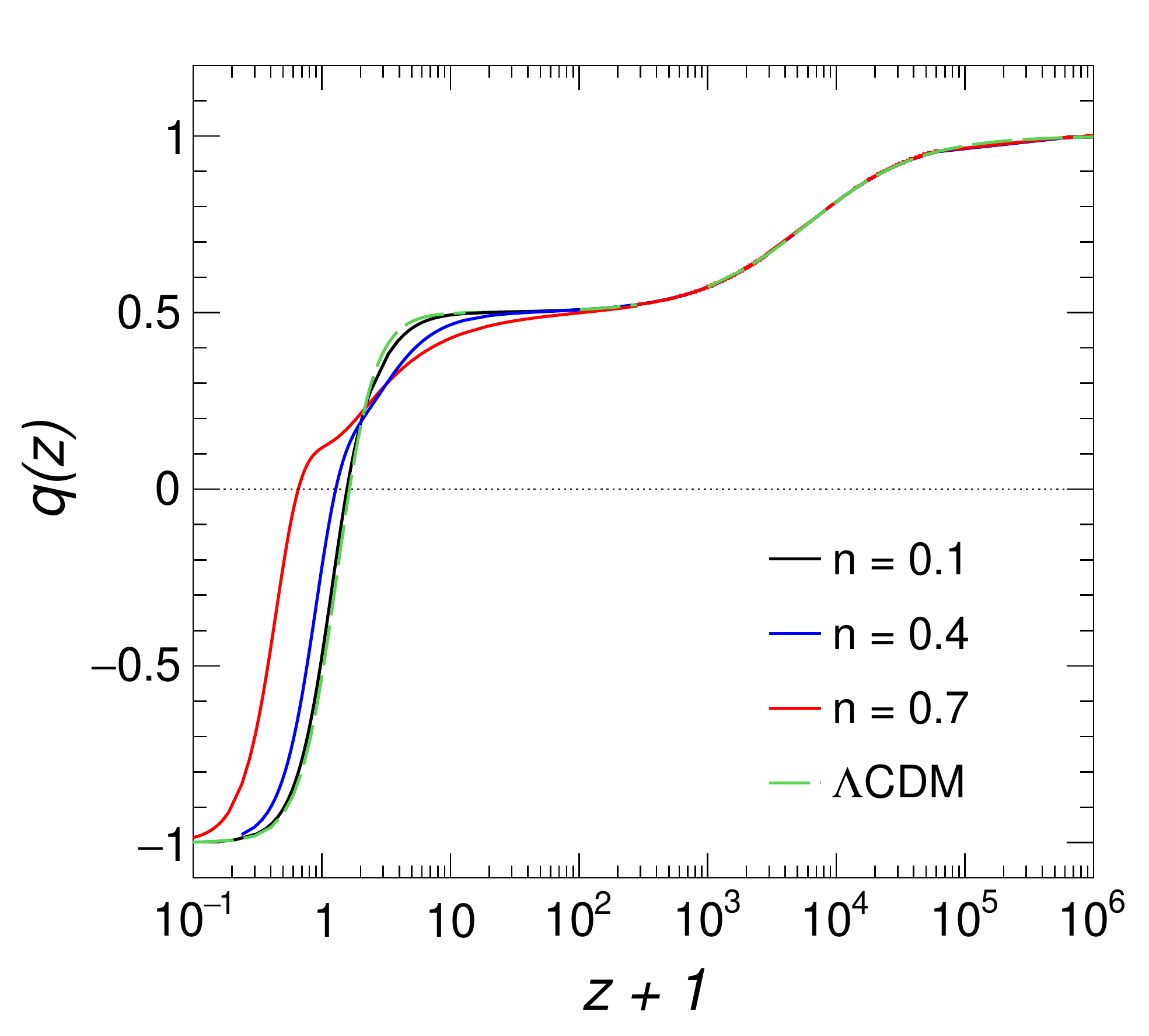}
}
\vspace{-0.2cm}
\caption{Pattern of variations of $\omega_{eff}(z)$ (left panel) and 
deceleration parameter $q(z)$ (right panel) with respect to $z$ for three 
constrained values of the parameter $n$ of the $f(R) = R + \alpha  H_0^{2 (1-n)} R^n$ model. The predictions of these two cosmological parameters by the 
$\Lambda CDM$ model are also shown in the corresponding plots.}
\label{fig06}
\end{figure}

On the left panel of Fig.\ \ref{fig06}, we have plotted the $\omega_{eff}(z)$ 
versus $z+1$ plot for this model. Unlike the power law model, for this case, 
we see that the model can mimic $\Lambda$CDM model in a better way. The 
deviations are again larger for $n=0.7$ from the $\Lambda$CDM model mostly 
within the region of small values of $z$. Around $n=0.1$, the model predicts 
the behaviour of the $\Lambda$CDM model in a comparatively better form. To be 
more specific, in this model the universe can have all the three phases of 
evolution similar to the $\Lambda$CDM model, i.e.\ the universe in this model 
starts from the radiation dominated phase, covers the matter dominated phase 
and finally ends with late time accelerating phase. These results are in a good 
agreement with the previous results of the model \cite{Cao2018}. 
Similar results are 
obtained for the deceleration parameter $q(z)$ versus redshift $z$ calculations
as predicted by the model (see the right plot of Fig.\ \ref{fig06}). 

\subsection{ The Gogoi-Goswami model}
\begin{figure}[htb]
\centerline{
   \includegraphics[scale = 0.3]{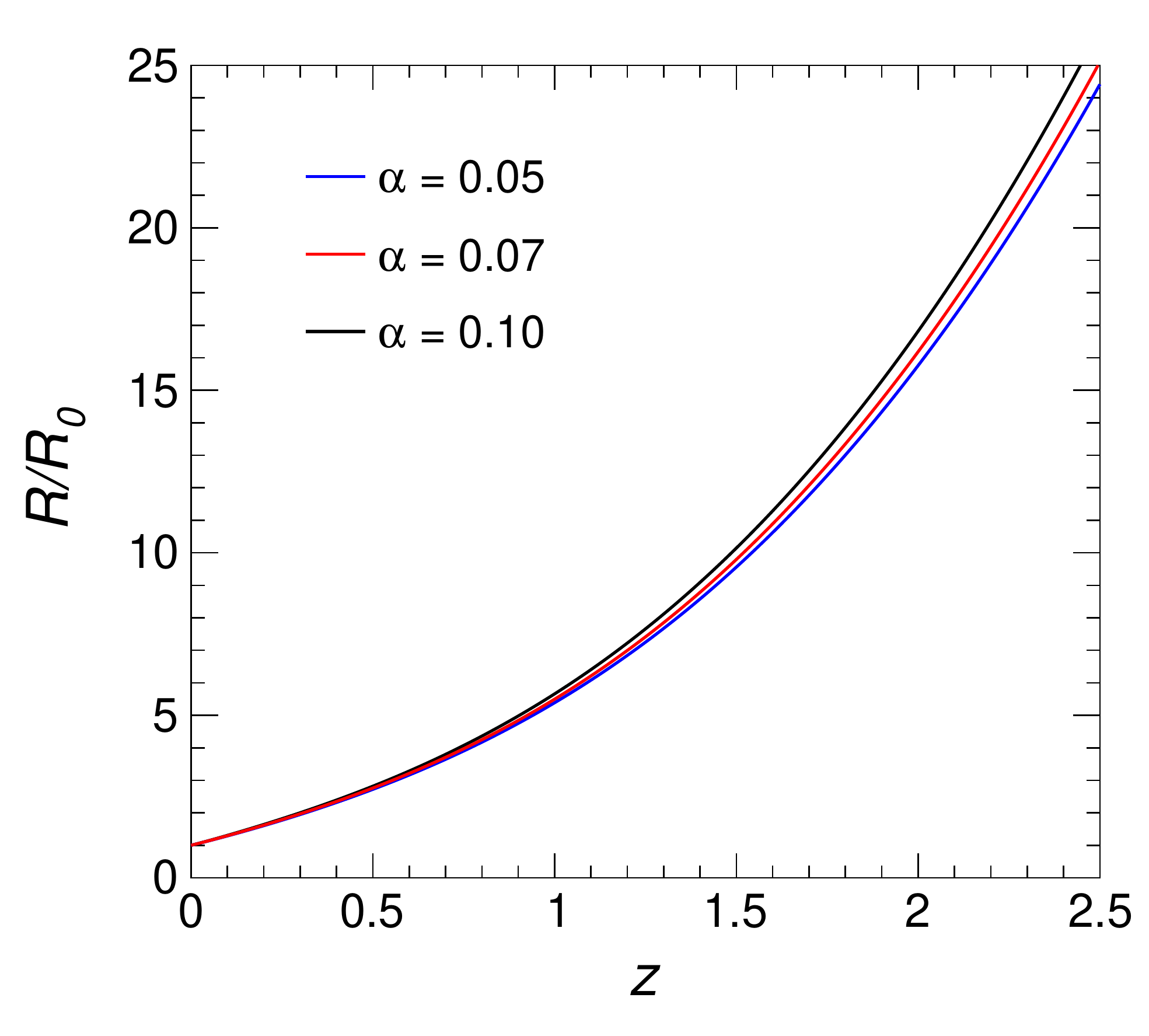}}
\vspace{-0.2cm}
\caption{Plot of $R/R_0$ versus $z$ for three different values of the parameter
$\alpha$ of the Gogoi-Goswami model.}
\label{fig07}
\end{figure}

Most recently introduced the Gogoi-Goswami model of $f(R)$ gravity is given 
by \cite{gogoi01}
\begin{equation}\label{GGmodel}
f(R) = R-\frac{\alpha}{\pi }\, R_c \cot ^{-1}\!\left(\tfrac{R_c^2}{R^2}\right)-\beta\,  R_c\! \left[1-\exp\left({-\,\tfrac{R}{R_c}}\right)\right],
\end{equation}
where $\alpha$ and $\beta$ are two dimensionless positive constants. $R_c$ 
is a characteristic curvature constant having dimensions same as curvature 
scalar $R$. This model was first defined in \cite{gogoi01}, where the model 
was studied in the metric formalism to see the properties of gravitational 
waves in it and the scalar degrees of freedom extensively. The model passes 
the solar system tests in the metric formalism and can mimic the $\Lambda$CDM 
model at large curvatures. Another important feature of this model is that it 
has two correction factors having different significant contributions to 
its behaviours as seen from the results studied in \cite{gogoi01}. Here, we 
shall focus on the Palatini formalism to see the cosmological implications of 
the model. For mathematical simplicity, we shall assume that the characteristic
curvature constant $R_c$ is equal to the background curvature of the present 
epoch. So, hereafter, we will use $R_c = R_0$. For this model, in the limit 
$z \rightarrow 0$, Eq.~\eqref{eq9} gives,
\begin{equation}\label{beta_GG}
\beta=-\frac{e \left(6\, \pi\,  H_0^2\, \Omega _{\text{m0}}+\pi  (\alpha -2) R_0-2 \alpha  R_0\right)}{2 (2 e-3) \pi  R_0},
\end{equation}
and Eq.\ \eqref{eq7} in the limit $z \rightarrow 0$ becomes,
\begin{equation}\label{H0_GG}
\frac{\pi  R_0^2 (e (\pi -2 \alpha )-2 \pi  \beta )^2 (e (\pi -\alpha )-\pi  \beta ) \left(12 e H_0^2 \left(\Omega _{\text{m0}}+2 \Omega _{\text{r0}}\right)-e R_0 (\alpha +4 \beta -4)+4 \beta  R_0\right)}{6 \left(2 H_0 R_0 (e (\pi -2 \alpha )-2 \pi  \beta ) (e (\pi -\alpha )-\pi  \beta )-9 e \pi  H_0^3 (e \alpha +\pi  \beta )\, \Omega _{\text{m0}}\right){}^2}=1.
\end{equation}
Using the value of $\beta$ given in the above expression we can solve for 
the present background curvature $R_0$ numerically. Now, $z-R$ relation can be 
easily obtained using the Eq.~\eqref{eq9}, which is
\begin{equation}\label{z_GG}
z=\frac{3^{2/3} \sqrt[3]{H_0^4\, \Omega _{\text{m0}}^2 \left(-2 \alpha  R_0 \cot ^{-1}\left(\frac{R_0^2}{R^2}\right)+\frac{2 \alpha  R^2 R_0^3}{R^4+R_0^4}+\pi  \beta  e^{-\frac{R}{R_0}} \left(R+2 R_0\right)+\pi  \left(R-2 \beta  R_0\right)\right)}-3 \sqrt[3]\,{\pi }\, H_0^2\, \Omega _{\text{m0}}}{3 \sqrt[3]\,{\pi }\, H_0^2\, \Omega _{\text{m0}}}.
\end{equation}
The above expression shows that it is difficult to obtain an analytical 
solution of the Ricci scalar $R$ in terms of $z$. Therefore, we will express 
the cosmographic parameters as functions of $R$ as mentioned earlier. The
numerically calculated $z-R$ relations for three different values of the model
parameter $\alpha = 0.05, 0.07$ and $0.10$ are shown in Fig.\ \ref{fig07}. 
These three values of $\alpha$ are taken from the region allowed by the 
parameter space analysis of the model \cite{gogoi01}. Although the differences 
of the curvatures of $R$ for these three values of $\alpha$ are very small, we 
see that the slope of the curvature is smaller for smaller values of $\alpha$,
similar to the case of the previously discussed Starobinsky type model.

\begin{figure}[htb]
\centerline{
   \includegraphics[scale = 0.3]{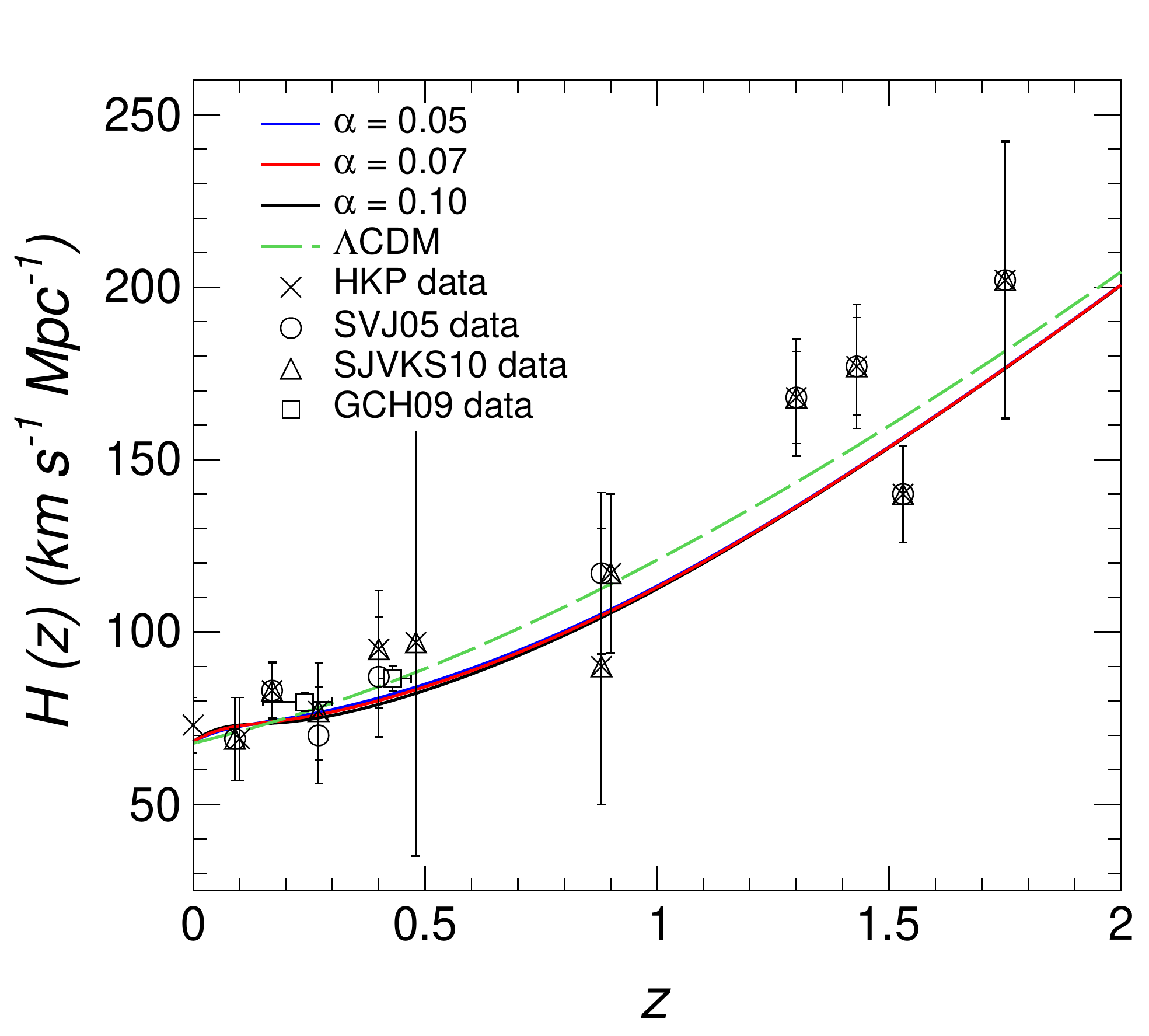}\hspace{0.5cm}
   \includegraphics[scale = 0.3]{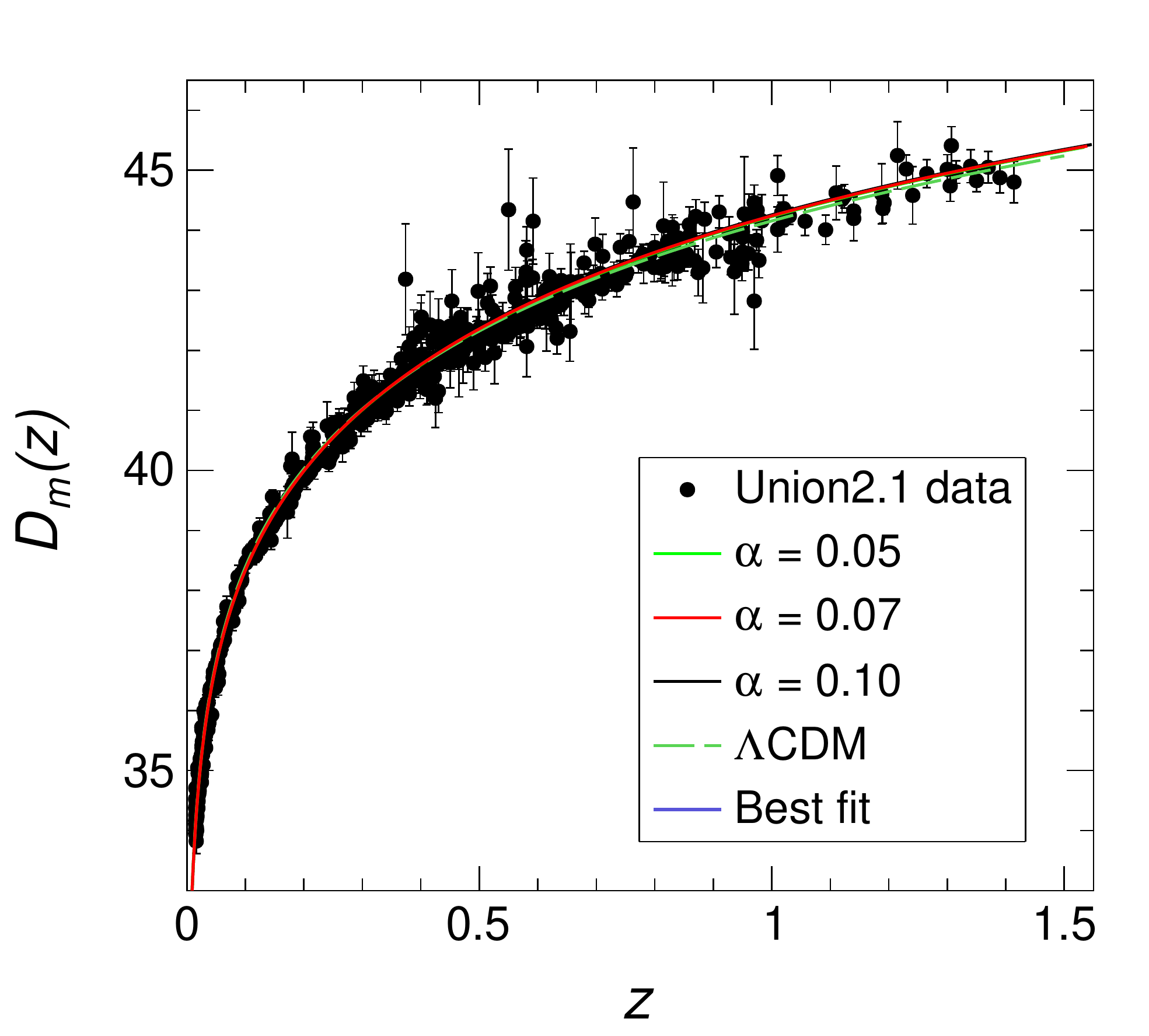}}
\vspace{-0.2cm}
\caption{Hubble parameter $H(z)$ versus $z$ (left) and distance modulus 
$D_m$ versus $z$ (right) plots for different values of parameter $\alpha$ 
of the Gogoi-Goswami model. The three values of $\alpha$ are taken from the
parameter space analysis of the model \cite{gogoi01} to fit with four sets of 
observed data in the $H(z)$ plot. These values of $\alpha$ are then used to 
fit the Union$2.1$ data \cite{Suzuki2012} in the $D_m(z)$ plot. As in the case
of other models, in both plots the corresponding $\Lambda CDM$ model 
predictions are also shown.}
\label{fig08}
\end{figure}

For this model, the expression for $H$ is given by
\begin{equation} \label{H_GG}
H = \frac{\frac{H_0}{\sqrt{-\frac{12 \alpha  R R_0^3}{\pi  R^4+\pi  R_0^4}-6 \beta  e^{-\frac{R}{R_0}}+6}} \sqrt{\frac{-\frac{\alpha  R_0 \cot ^{-1}\left(\frac{R_0^2}{R^2}\right)}{\pi }+\beta  \left(e^{-\frac{R}{R_0}}-1\right) R_0+R}{H_0^2}+3 (z+1)^3 \Omega _{\text{m0}}+6 (z+1)^4 \Omega _{\text{r0}}}}{ 1-\frac{9 \pi  H_0^2 e^{\frac{R}{R_0}} \left(R^4+R_0^4\right) (z+1)^3 \Omega _{\text{m0}} \left(\pi  \beta  \left(R^4+R_0^4\right){}^2-2 \alpha  e^{\frac{R}{R_0}} R_0^4 \left(R_0^4-3 R^4\right)\right)}{2 \left(e^{\frac{R}{R_0}} \left(\pi  \left(R^4+R_0^4\right)-2 \alpha  R R_0^3\right)-\pi  \beta  \left(R^4+R_0^4\right)\right) \left(e^{\frac{R}{R_0}} R_0 \left(\pi  \left(R^4+R_0^4\right){}^2-8 \alpha  R^5 R_0^3\right)-\pi  \beta  \left(R+R_0\right) \left(R^4+R_0^4\right){}^2\right)}}.
\end{equation}
Now, using Eq.\ \eqref{z_GG} in \eqref{H_GG} we can obtain $H$ as a function 
of $R$ instead of $z$. With an intention similar to the case of other two 
models, on the left panel of Fig.\ \ref{fig08}  
we have plotted $H(z)$ versus $z$ for this model with the model parameter 
$\alpha = 0.05, 0.07$ and $0.10$ as used for the $z-R$ plot and compared with 
the experimental HKP data, SVJ05 data, SJVKS10 data and GCH09 data. In this 
case also, the model shows good behaviour with the experimental data as well
as with the prediction of the $\Lambda$CDM model. It is seen that the 
differences in $z-H$ curves for all three used values of $\alpha$ are almost 
negligible. However, the values of $H(z)$ for $\alpha = 0.05$ are slightly 
greater than those corresponding to $\alpha = 0.07$ and $0.10$, especially in 
the region $z\le 1$. 

Numerically calculated values of the distance modulus $D_m(z)$ from the 
Eq.\ \eqref{Dm} for the Gogoi-Goswami model for the said values of the 
parameter $\alpha$ are shown on the right panel of Fig.\ \ref{fig08}. It is 
seen that all these three values of $\alpha$ show a good agreement with the 
observational Union2.1 data and with the $\Lambda$CDM prediction as expected 
from the previous two plots. This shows a better suitability of the parameter 
of the model to the experimental data in comparison to other two models.

\begin{figure}[htb]
\centerline{
   \includegraphics[scale = 0.3]{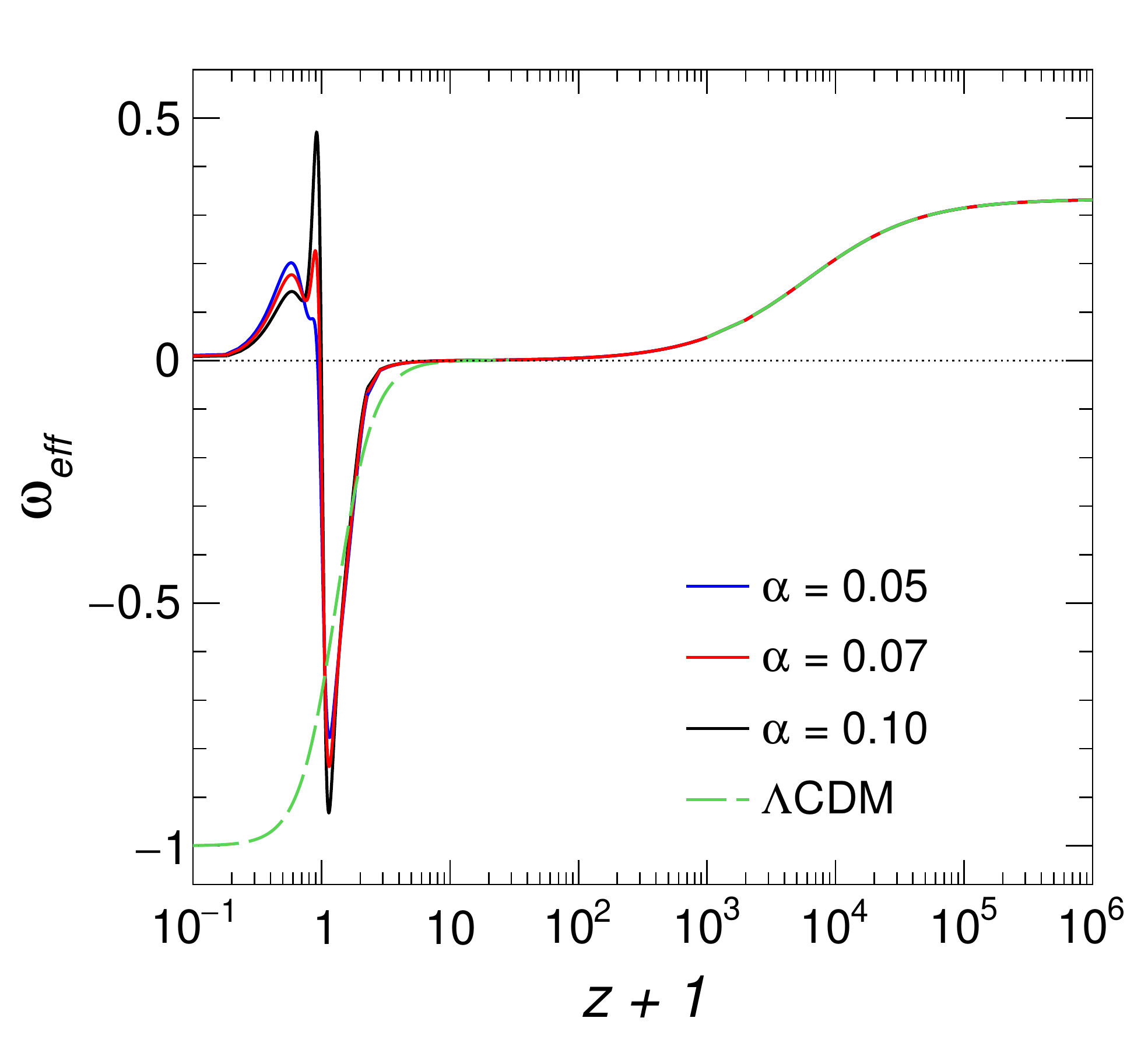}\hspace{0.5cm}
   \includegraphics[scale = 0.3]{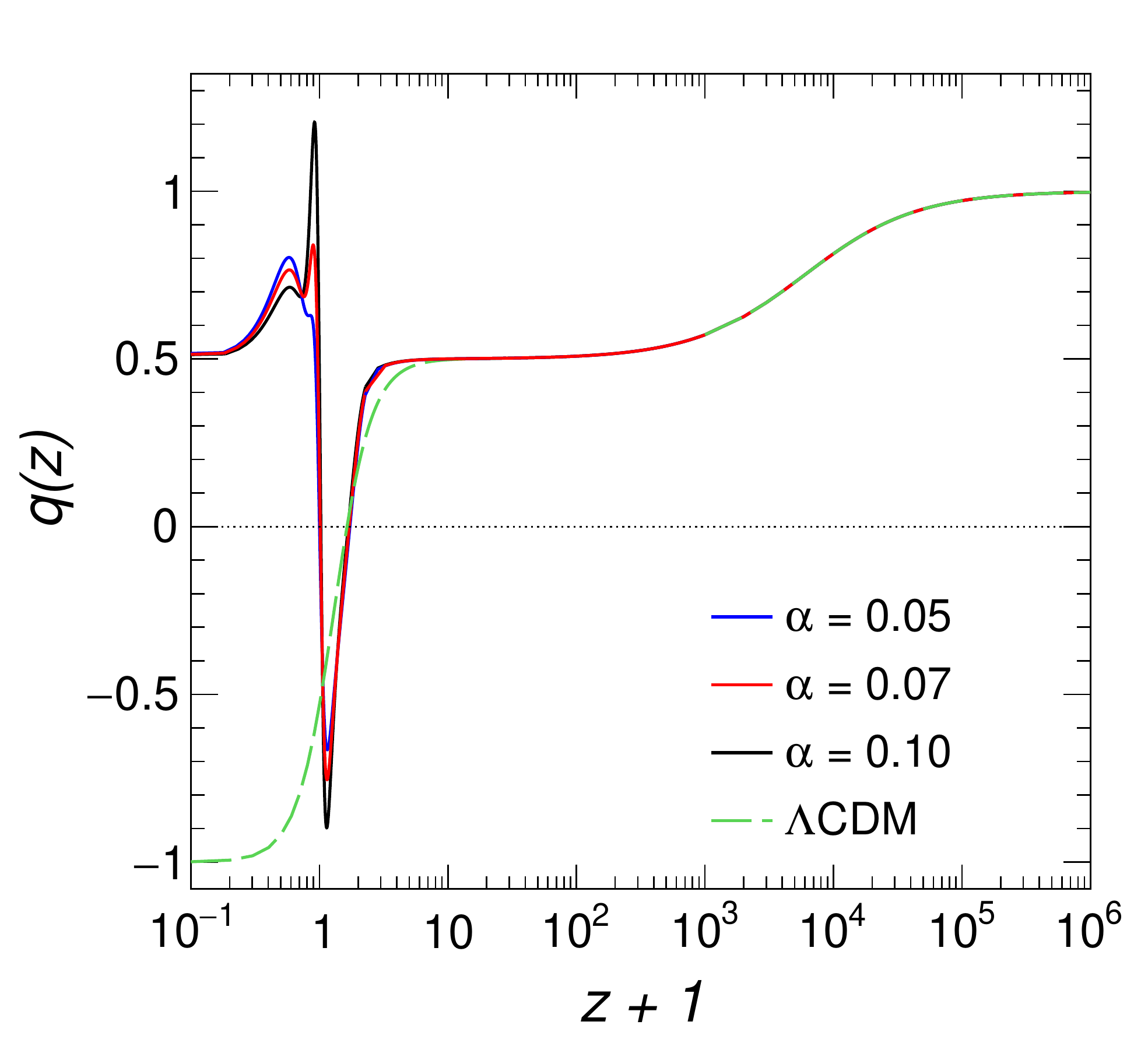}
}
\vspace{-0.2cm}
\caption{$\omega_{eff}(z)$ (left) and deceleration parameter $q(z)$ (right)
variations with respect to redshift $z$ obtained for the Gogoi-Goswami model
for three different values of the model parameter $\alpha$. As in the previous
cases the predictions of these two cosmological parameters by the $\Lambda CDM$
model are also shown in the corresponding plots.}
\label{fig09}
\end{figure}

Variations of $\omega_{eff}(z)$ and $q(z)$ with respect to $z+1$ for the 
Gogoi-Goswami model are shown in the left and right plots respectively of 
Fig.\ \ref{fig09} for the same set of values of $\alpha$ used above. We see 
that in the near future at around $z=-1$, $\omega_{eff}$ approaches zero 
for this model showing large deviations from the $\Lambda$CDM model. However, 
in the present universe, the model mimics the results of $\Lambda$CDM model 
and then with a slight deviation it follows the trend of the $\Lambda$CDM 
model towards the past. Finally in the early universe, the model behaviour is 
totally identical to the $\Lambda$CDM model. That is, in the early time, 
the universe in this model starts from the radiation dominated phase and then 
as time passes it attains the matter dominated phase in exactly the same way of 
the $\Lambda$CDM model. Towards the present time, the universe shows late time 
accelerating phase with a slight deviation from the $\Lambda$CDM model and in 
the near future, the universe is crossing the matter dominated phase towards 
the radiation dominated phase and finally ends near the matter dominated phase.
This part of future time ($z<0$ region) behaviour of the model is unique and 
does not match with the evolution phase of the universe during this period 
in the $\Lambda$CDM model. Similar results hold for the deceleration parameter 
$q(z)$ versus redshift $z$ plot also as seen from the right panel of 
Fig.\ \ref{fig09}.

\section{Constraints on the Models from Observed Hubble Data} \label{sec05}
Although in the last section we have used four sets of Hubble parameter data
to constrained the parameters of the models within a possible range of viable
parameter space, in this section we use all possible Observed Hubble Data (OHD)
available to us, shown in the Table \ref{Hubble} to get a knowledge on 
the feasible parameter space by constraining the models in a comprehensive way.
This will enable us to see or predict the behaviour of the model for the 
reliable parameter allowed or constrained by OHD. Here, we shall use the OHD 
dataset to fit the models considered in this study and to see the goodness of 
the fitting we shall implement the $\chi^2$ minimization method defined as
\begin{equation}\label{chi}
  \chi ^2=\sum_i\frac{[H_{\rm th}(z_i|\textbf{p})-H_{\rm obs}(z_i)]^2}{\sigma^2(z_i)},
\end{equation}
where $H_{\rm th}(z_i|\textbf{p})$ is the theoretical value of the Hubble 
parameter $H$ at a specific redshift $z_i$ and for parameters $\textbf{p}$ 
depend on the $f(R)$ models, $H_{\rm obs}(z_i)$ are the OHD, and $\sigma(z_i)$ 
is the uncertainty of each $H_{\rm obs}(z_i)$ as obtained in OHD. Here we 
assume that each measurement in $H_{\rm obs}(z_i)$ is independent.

\begin{figure}[htb]
\centerline{
   \includegraphics[scale = 0.35]{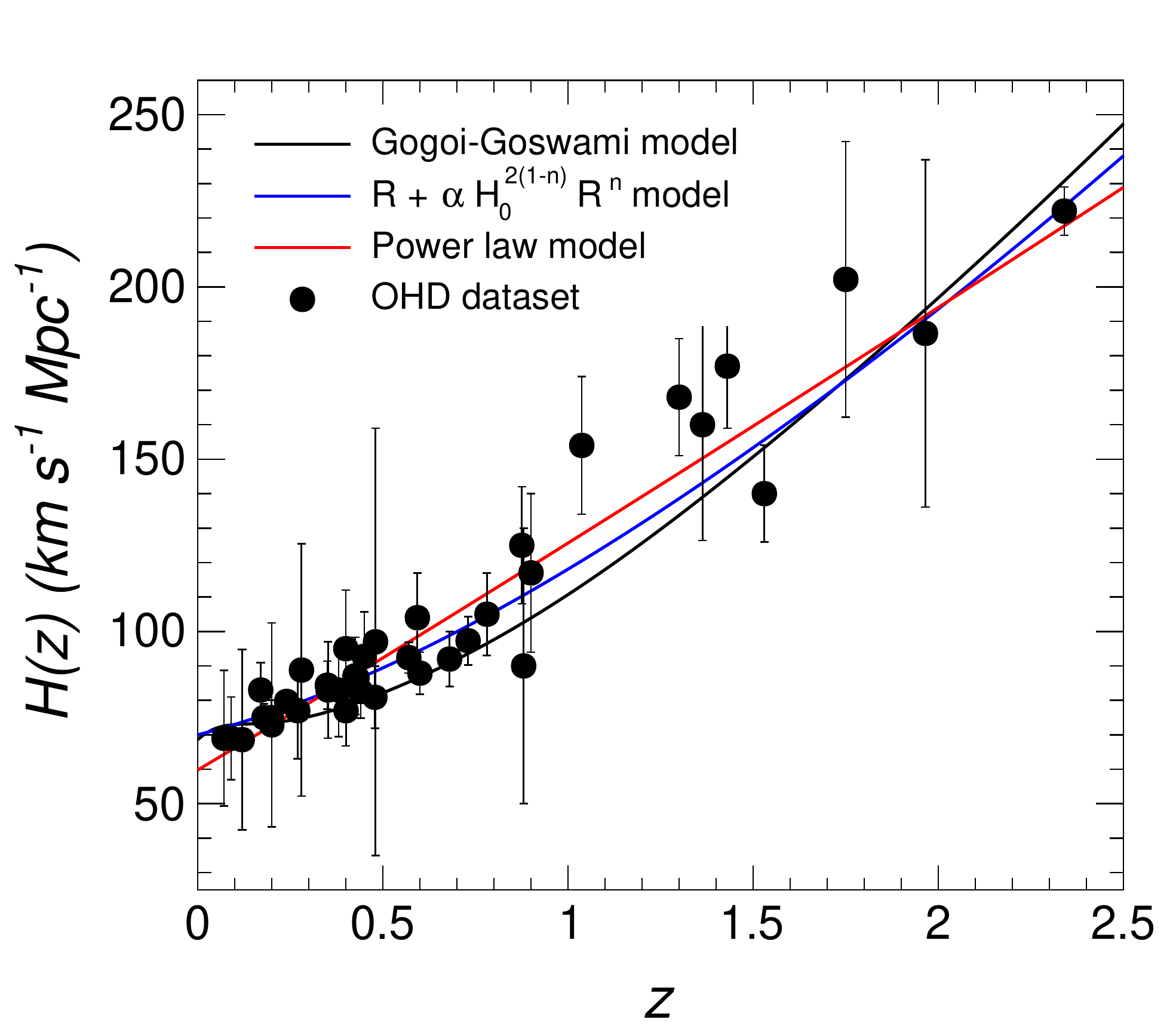}}
\vspace{-0.2cm}
\caption{$\chi^2$ fitting to the OHD dataset obtained for the power law model, 
$R + \alpha  H_0^{2 (1-n)} R^n$ model and Gogoi-Goswami model with the model 
parameters $n = 1.4$, $n = 0.1$ and $\alpha = 0.1$ respectively. In this 
fitting the respective $\chi^2$ values are found as $27.564$, $18.303$ and 
$24.864$. }
\label{fig10}
\end{figure}

To make the expressions comparatively simple for the calculation of 
$H_{\rm th}(z_i|\textbf{p})$ values for the models, we have neglected the 
radiation component contribution in the OHD as the associated values of 
$z$ are not very high. After this simplification, for the power law model one 
can see from the previous section that the Hubble parameter becomes independent
of the term $\Omega_{m0}$. So the set of parameters considered for this model 
are: $\textbf{p}=(H_0,n)$. And from the best fit, we have obtained for the 
power law model that $n=1.4$ and $H_0 = 59.8\ \rm km/s/Mpc$. This best value 
of $n$ almost agrees with the earlier best fit value of $n = 1.38$ used in the 
previous section for this model. For the model 
$f(R) = R + \alpha  H_0^{2 (1-n)} R^n$, the set of parameters considered 
are: $\textbf{p}=(H_0,n, \Omega_{m0})$. In this case, we have obtained the 
best fit values as $n=0.10$, $\Omega_{m0}=0.24$ and $H_0=70\ \rm km/s/Mpc$. 
The values of $\Omega_{m0}$ and $H_0$ found in this best fit are close to the 
results obtained in \cite{Amarzguioui2006, Cao2018}. Similarly, for the 
Gogoi-Goswami model, we have considered $\textbf{p}=(H_0,\alpha, \Omega_{m0})$ 
and the best fit values are found as $\alpha=0.1$, $\Omega_{m0}=0.30$ and 
$H_0=68.6\ \rm km/s/Mpc$. Here the best fitted values of $\Omega_{m0}$ 
and $H_0$ are relatively near to their values found in \cite{Zhang2021} 
and in recent Planck's results \cite{Planck2018}. Fig.\ \ref{fig10} shows the
best fit plots to the OHD for the all three models.     

\begin{table}
\centering
\begin{tabular}{|llcc|llcc|}
\hline
{$\;\;\;z$} & $\;\;\;\;H(z)$ & Method & Reference& {$\:\:\;z$} & $\;\;\;\;H(z)$ & Method & Reference\\
\hline
$0.0708$ & $69.0\pm19.68$ & DGAM & \cite{Zhang2014} & $0.4783$ & $80.9\pm9.0$ & DGAM & \cite{Moresco2016}   \\
$0.09$ & $69.0\pm12.0$ &  DGAM &  \cite{Jimenez2003} & $0.48$ & $97.0\pm62.0$ &  DGAM & \cite{Stern2010}  \\
$0.12$ & $68.6\pm26.2$ & DGAM & \cite{Zhang2014} & $0.57$ & $92.4\pm4.5$ &  RBAOM & \cite{Samushia2013}  \\
$0.17$ &  $83.0\pm8.0$          &   DGAM    &  \cite{Simon2005} &  $0.593$     &  $104.0\pm13.0$      &   DGAM    &  \cite{Moresco2012} \\
    $0.179$     &  $75.0\pm4.0$          &   DGAM    &  \cite{Moresco2012} & $0.6$         &  $87.9\pm6.1$          &  RBAOM   &  \cite{Blake2012}    \\
    $0.199$     &  $75.0\pm5.0$          &   DGAM    &  \cite{Moresco2012} &  $0.68$       &  $92.0\pm8.0$          &   DGAM    &  \cite{Moresco2012}    \\
    $0.20$         &  $72.9\pm29.6$        &   DGAM    &  \cite{Zhang2014} & $0.73$       &  $97.3\pm7.0$          &  RBAOM   &  \cite{Blake2012} \\
    $0.240$     &  $79.69\pm2.65$      &  RBAOM   &  \cite{Gaztanaga2009} &  $0.781$     &  $105.0\pm12.0$      &   DGAM   &  \cite{Moresco2012}  \\
    $0.27$       &  $77.0\pm14.0$        &   DGAM    &    \cite{Simon2005} & $0.875$     &  $125.0\pm17.0$      &   DGAM    &  \cite{Moresco2012} \\
    $0.28$       &  $88.8\pm36.6$        &   DGAM    &  \cite{Zhang2014} &  $0.88$       &  $90.0\pm40.0$        &   DGAM   &  \cite{Stern2010}   \\
    $0.35$       &  $84.4\pm7.0$          &  RBAOM   &   \cite{Xu2013} &  $0.9$         &  $117.0\pm23.0$      &   DGAM    &  \cite{Simon2005} \\
    $0.352$     &  $83.0\pm14.0$        &   DGAM   &  \cite{Moresco2012} & $1.037$     &  $154.0\pm20.0$      &   DGAM    &  \cite{Moresco2012} \\
    $0.3802$     &  $83.0\pm13.5$        &   DGAM    &  \cite{Moresco2016} & $1.3$         &  $168.0\pm17.0$      &   DGAM    &  \cite{Simon2005}  \\
    $0.4$         &  $95\pm17.0$           &   DGAM    &  \cite{Simon2005}  & $1.363$     &  $160.0\pm33.6$      &   DGAM   &  \cite{Moresco2015}  \\
    $0.4004$     &  $77.0\pm10.2$        &   DGAM    &  \cite{Moresco2016} & $1.43$       &  $177.0\pm18.0$      &   DGAM    &  \cite{Simon2005} \\
    $0.4247$     &  $87.1\pm11.2$        &   DGAM    &  \cite{Moresco2016} & $1.53$       &  $140.0\pm14.0$      &   DGAM    &  \cite{Simon2005}   \\
    $0.43$     &  $86.45\pm3.68$        &  RBAOM   &  \cite{Gaztanaga2009}  & $1.75$       &  $202.0\pm40.0$      &    DGAM   &  \cite{Simon2005}   \\
    $0.44$       & $82.6\pm7.8$           &  RBAOM   &  \cite{Blake2012} &  $1.965$     &  $186.5\pm50.4$      &   DGAM    &   \cite{Moresco2015}  \\
    $0.4497$     &  $92.8\pm12.9$        &   DGAM    &  \cite{Moresco2016} & $2.34$       &  $222.0\pm7.0$        &  RBAOM   &  \cite{Delubac2015} \\
\hline
\end{tabular}
\caption{\label{Hubble} Currently available observed Hubble dataset. Here 
$H(z)$ is in units of ${\rm km/s/Mpc}$, DGAM stands for ``Differential Galactic 
Ages Method" and RBAOM for ``Radial BAO Method".} 
\end{table}

\section{Diagnostics of Models}\label{sec06}
In this section we analyze how much our models' cosmological behaviours are
different from the $\Lambda$CDM model as well as from each other by using two 
most effective diagnostics analyses, which are the $Om(z)$ test and 
statefinder diagnostic as follows.
\subsection{$Om(z)$ diagnostic} 
\begin{figure}[htb]
\centerline{
   \includegraphics[scale = 0.28]{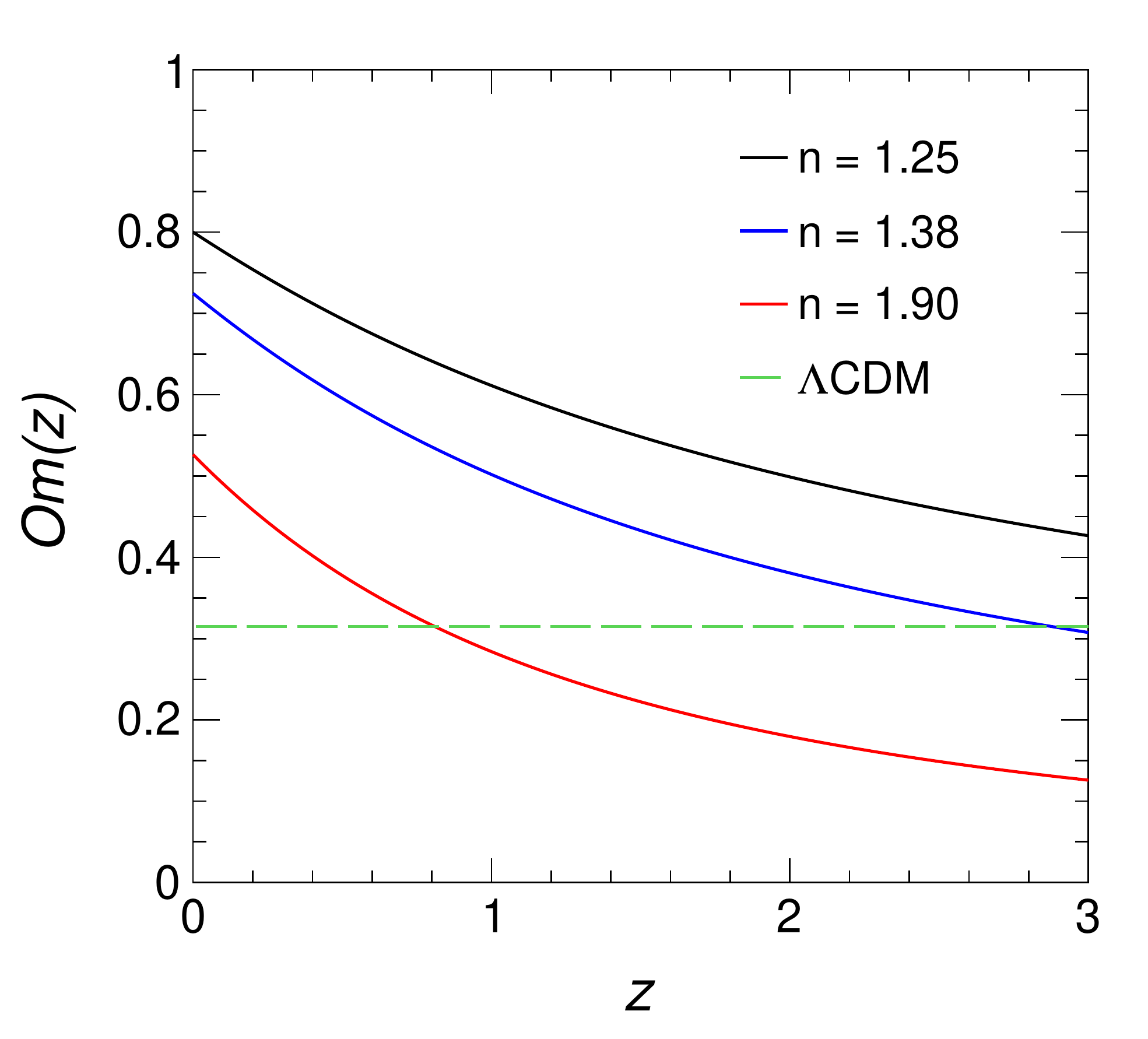}\hspace{0.1cm}
   \includegraphics[scale = 0.28]{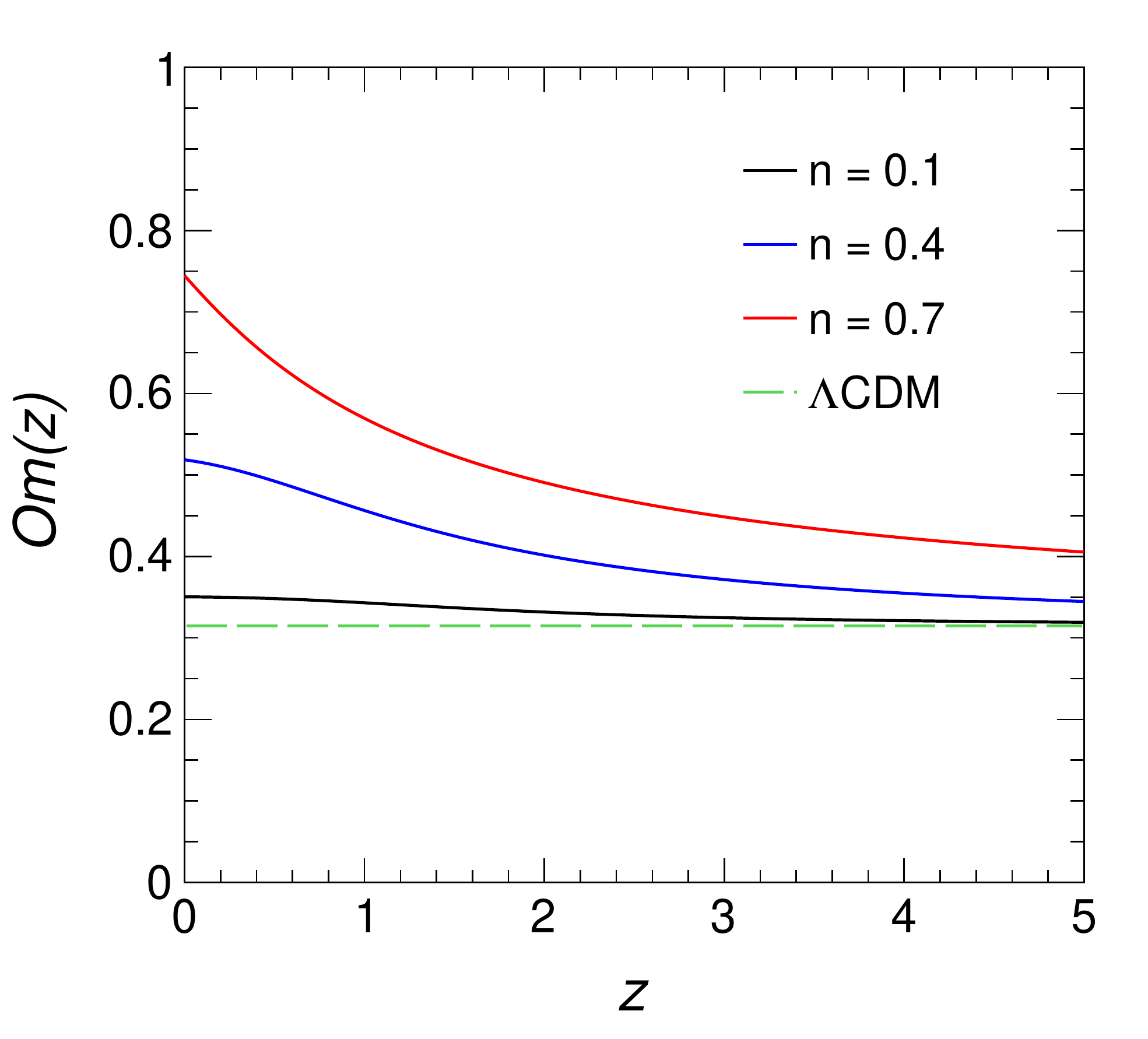}\hspace{0.1cm}
   \includegraphics[scale = 0.28]{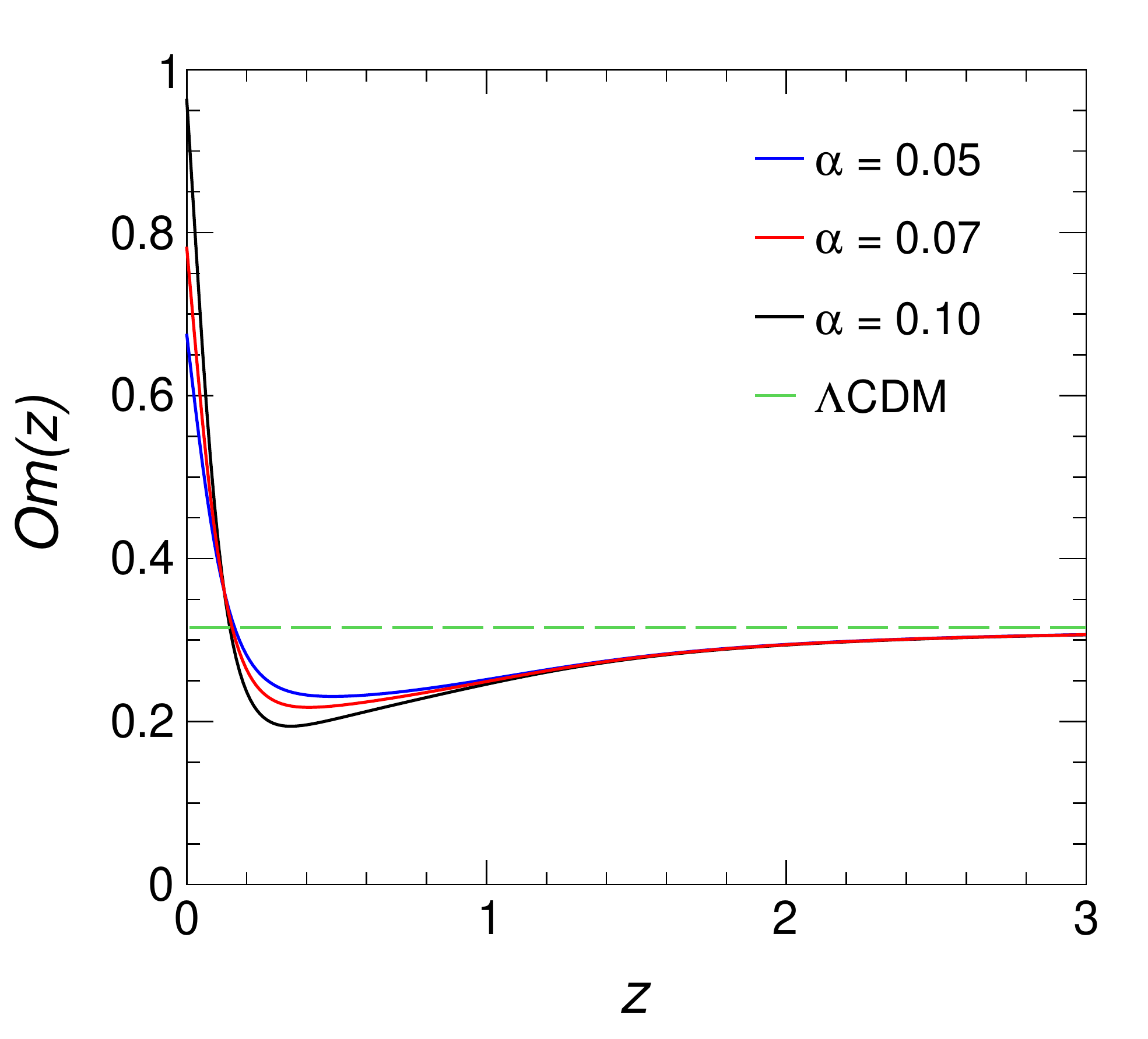}}
\vspace{-0.2cm}
\caption{$Om(z)$ diagnostics for power law model (left), 
$f(R) = R + \alpha  H_0^{2 (1-n)} R^n$ model (middle) and the Gogoi-Goswami 
model (right).}
\label{fig11}
\end{figure}

The $Om(z)$ diagnostic is a kind of test, which can be used to discriminate 
between different cosmological or dark energy models from the $\Lambda$CDM 
model. This diagnostic was first introduced in \cite{Om_diagnostic} and 
subsequently used extensively by different authors for the said purpose. This 
diagnostic parameter is defined by
 \begin{equation}\label{Omz} 
 Om(z)=\frac{E^2(z)-1}{(1+z)^3-1},
\end{equation}
where $E(z)=H(z)/H_0$ is the dimensionless Hubble expansion rate. Being 
negligible, if we ignore the radiation component at very low redshift, we can 
write the Friedmann equation for the $\Lambda$CDM model as
\begin{equation}
H(z)^2=H_0 ^2\left[\Omega_{\rm
m}(1+z)^3+1-\Omega_{\rm m}\right].
\end{equation}
This shows that if any cosmological model behaves exactly as the $\Lambda$CDM 
model, then the $Om(z)$ parameter of that model should be exactly equal to 
$\Omega_m$. Thus, this diagnostic can be used to compare and explicitly 
illustrate the difference between different $f(R)$ gravity cosmological models 
and the $\Lambda$CDM model. For the $f(R)$ gravity models considered in this 
paper, we have plotted the evolution of $Om(z)$ with respect to $z$ in 
Fig.\ \ref{fig11}. For the power law model, the $Om(z)$ function shows large 
deviations in the present universe and it approaches the $\Lambda$CDM model 
in the early universe. However, it is seen that the model can not completely 
mimic the $\Lambda$CDM model for a long time in the early universe also, as 
mimicking is only momentary in nature. The model parameter $n=1.90$ coincides 
the $\Lambda$CDM model just before $z=1$ and then starts deviating from it. 
This coinciding point moves towards the higher $z$ values with the decreasing 
value of $n$ (see left plot of Fig.\ \ref{fig11}).

For the model of the type $f(R) = R + \alpha  H_0^{2 (1-n)} R^n$, $n=0.1$ 
shows the smallest deviations from the $\Lambda$CDM model and starts mimicking 
it in the early universe. However, $n=0.4$ and $n=0.7$ show comparatively 
large deviations from the $\Lambda$CDM model in the present universe, which 
decreases towards the early universe. So, in conclusion, we can say that 
this model can be indistinguishable from the $\Lambda$CDM model practically 
in the early universe for smaller values of $n$ (see middle plot of 
Fig.\ \ref{fig11}).

On the right panel of Fig.\ \ref{fig11}, it is seen that the Gogoi-Goswami 
model is practically indistinguishable from the $\Lambda$CDM model in the 
early universe for all allowed values of the parameter $\alpha$. Although 
the model parameter $\alpha=0.05$ shows closer results to the $\Lambda$CDM 
model in the near past scenario, all the three values of it are seen to be 
indistinguishable in the early universe. As a whole, the $Om(z)$ function 
sharply decreases in the low redshift regime and crosses the $\Lambda$CDM 
model. After that, it starts following the $\Lambda$CDM model towards higher 
redshifts and becomes indistinguishable towards the early universe.

It is to be noted that these results are in agreement with previous results 
obtained from the behaviours $\omega_{eff}(z)$ and $q(z)$ with respect to $z$ 
for all three $f(R)$ gravity models.


\subsection{Statefinder diagnostic}
From the previous results, it is seen that the parameters $H(z)$, 
$\omega_{eff}(z)$ and $q(z)$ are not able to differentiate between 
different $f(R)$ gravity cosmological models effectively. One can see that 
$H(z)$ and $q(z)$ are respectively related to $\dot{a}$ 
and $\ddot{a}$. Thus in order to differentiate between different cosmological 
models one might need new parameters, which are functions of higher order 
derivatives of the scale factor $a(t)$. Such an effective pair of parameters 
are the so-called statefinder parameters $\{r,s\}$ \cite{Sahni2003n, Alam2003n}. 
These parameters are defined as \cite{Sahni2003n, Alam2003n,Pasqua2017, Xu2018}
\begin{equation}
r=\frac{\dddot{a}}{aH^3}, \quad s=\frac{r-1}{3(q-1/2)},
\end{equation}
where the deceleration parameter $q(z)$ can be rewritten using Eq.\ 
\eqref{eq9b} as
\begin{equation}
q(z)=\frac{E' (z)}{E(z)}(1+z)-1
\end{equation}
with $E' (z) \equiv \mathrm{d} E(z)/\mathrm{d} z$. Using this equation it 
is possible to rewrite the statefinder parameters in the following way:
\begin{eqnarray}
\nonumber r(z)&=&1-2\frac{E^\prime
(z)}{E(z)}(1+z)+\left[\frac{E^{\prime
\prime}(z)}{E(z)}+\left(\frac{E^\prime
(z)}{E(z)}\right)^2\right](1+z)^2
\\
 &=&q(z)(1+2q(z))+q^\prime (z)(1+z),
\\
 s(z)&=&\frac{r(z)-1}{3(q(z)-1/2)},
\end{eqnarray}
where $q^\prime (z) \equiv \mathrm{d} q(z)/\mathrm{d} z$. Using the above 
equations we have calculated the parameters $r(z)$ and $s(z)$ for our set of
models with their respective sets of model parameters used earlier.

\begin{figure}[htb]
\centerline{
   \includegraphics[scale = 0.28]{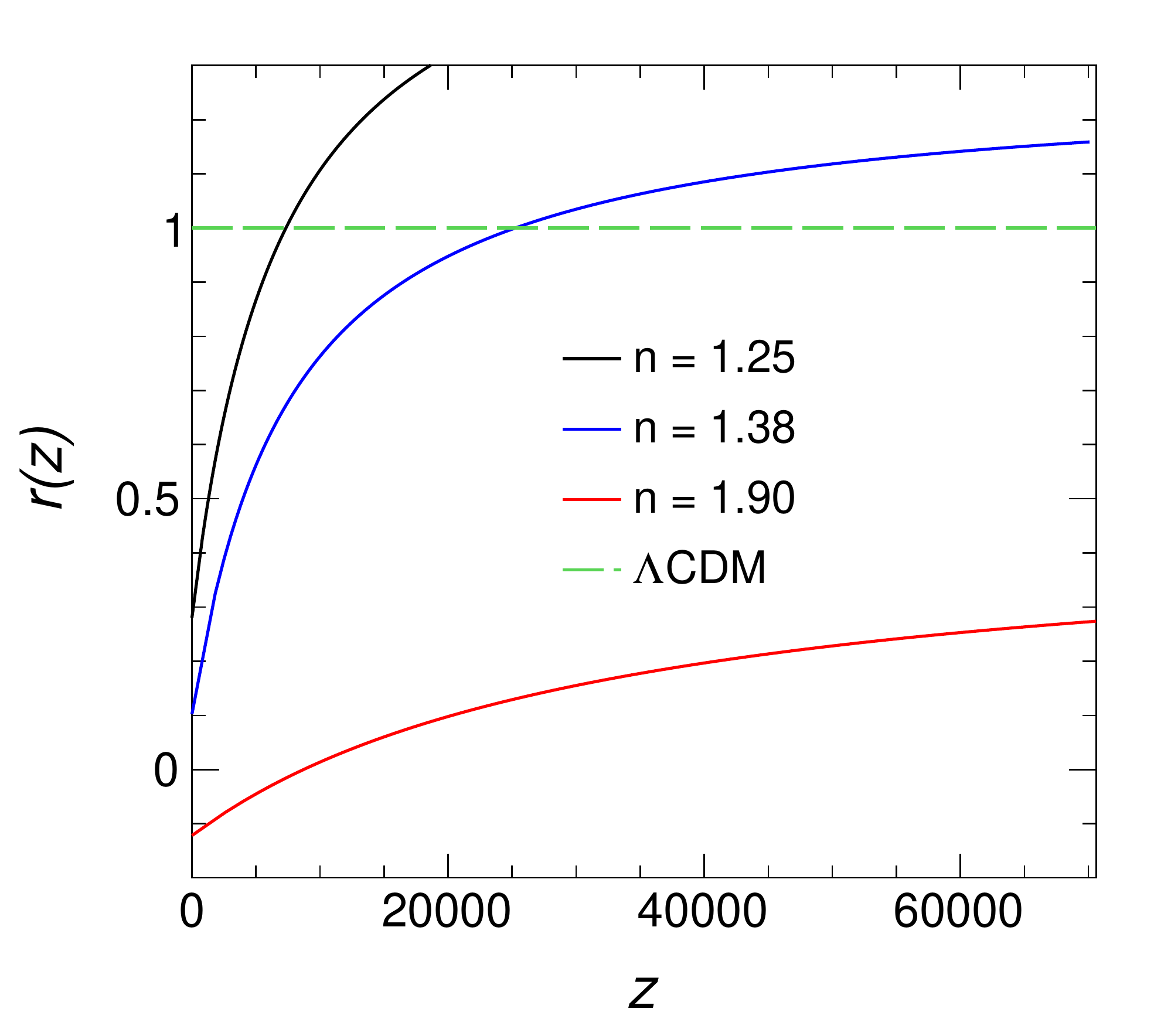}\hspace{0.2cm}
   \includegraphics[scale = 0.28]{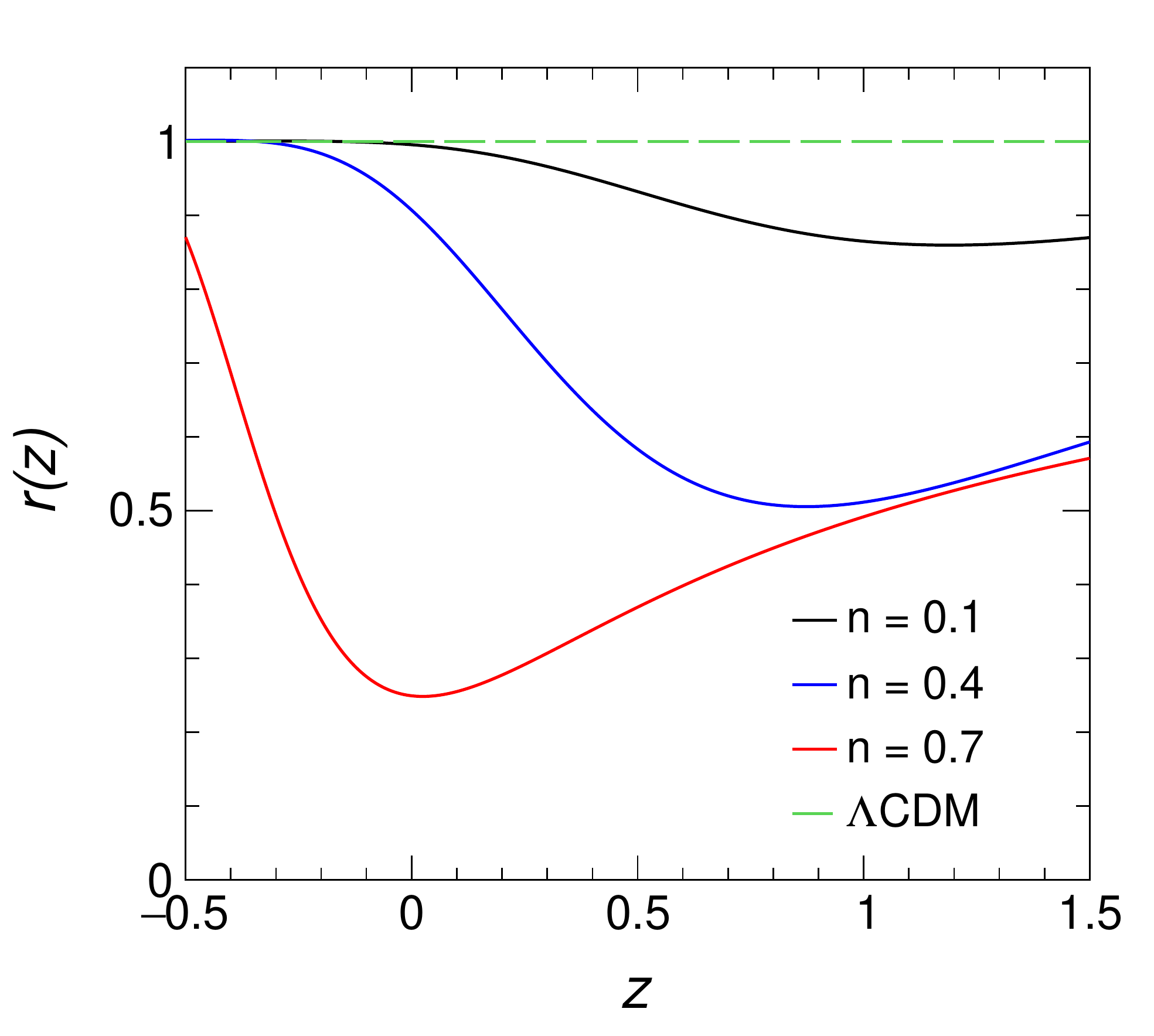}\hspace{0.2cm}
   \includegraphics[scale = 0.28]{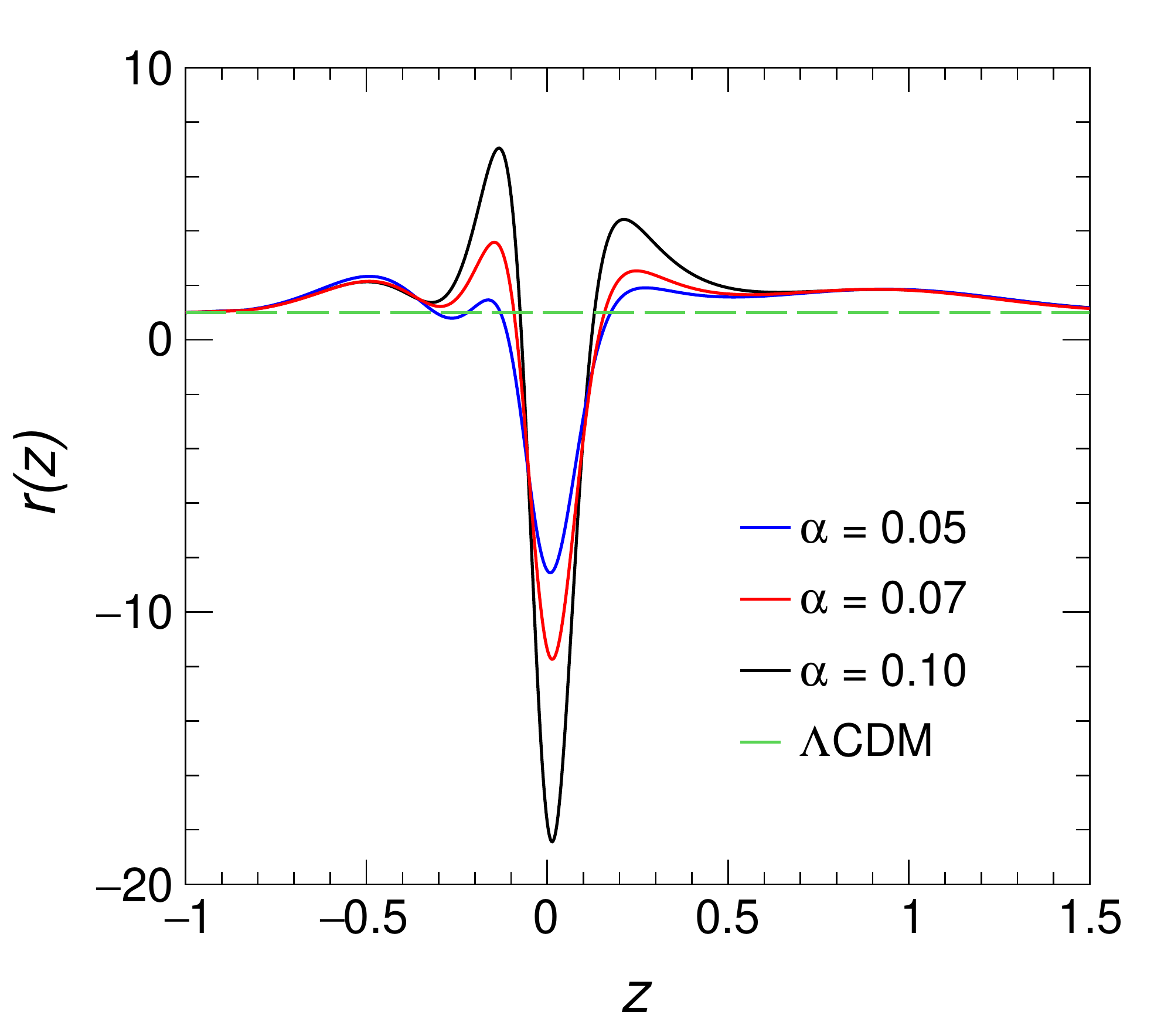}
}
\vspace{-0.2cm}
\caption{Statefinder parameter $r$ versus $z$ for the $f(R)$ gravity power law 
model (left), $f(R) = R + \alpha  H_0^{2 (1-n)} R^n$ model (middle) and 
Gogoi-Goswami model (right).}
\label{fig12}
\end{figure}
\begin{figure}[htb]
\centerline{
   \includegraphics[scale = 0.6]{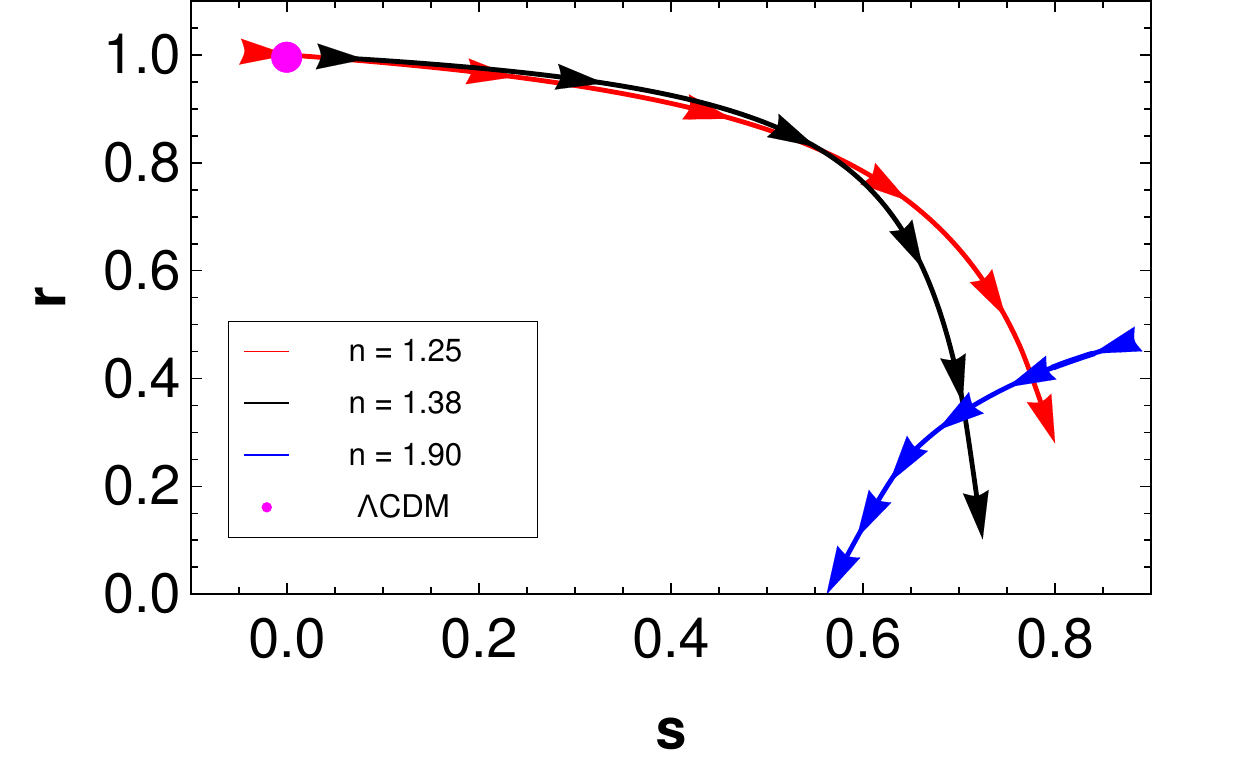}\hspace{0.5cm}
   \includegraphics[scale = 0.6]{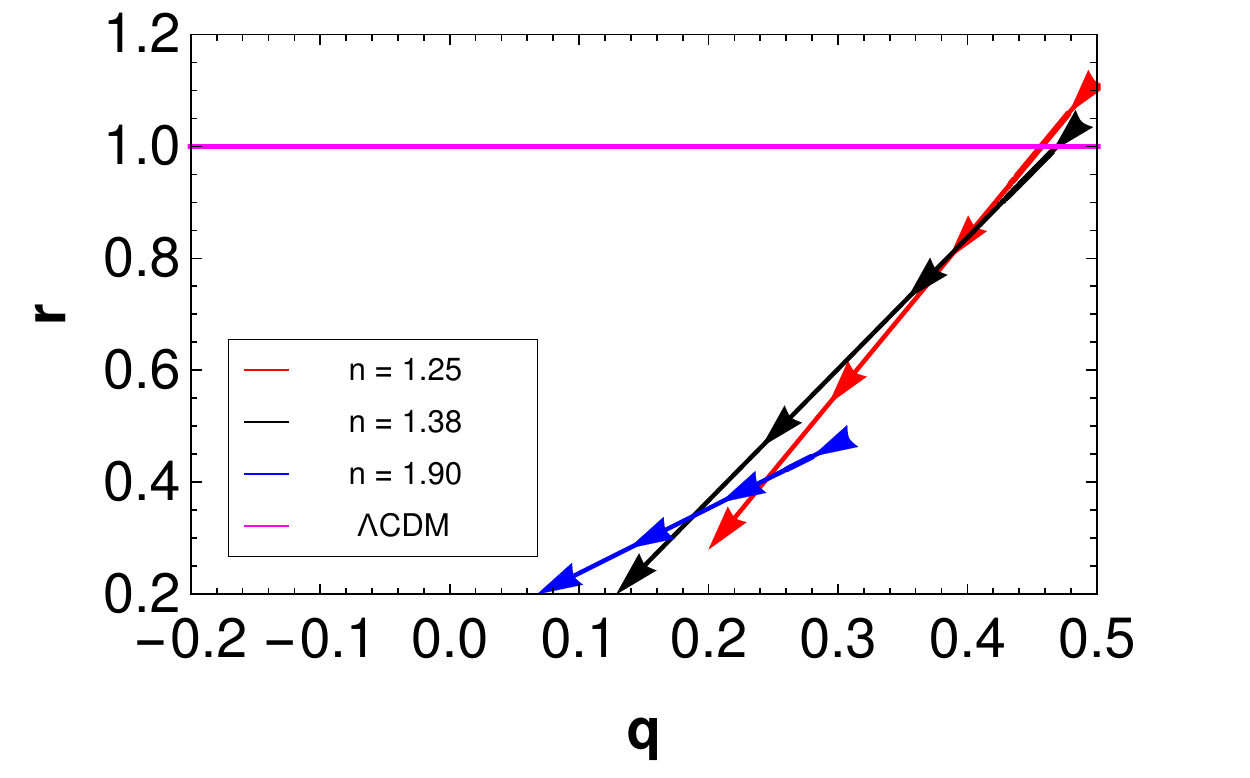}}
\vspace{-0.2cm}
\caption{Evolution of the statefinder parameter $r$ with respect to the 
parameter $s$ (left) and evolution of $r$ with the deceleration parameter $q$ 
(right) for the $f(R)$ gravity power law model.}
\label{fig13}
\end{figure}

\begin{figure}[htb]
\centerline{
   \includegraphics[scale = 0.6]{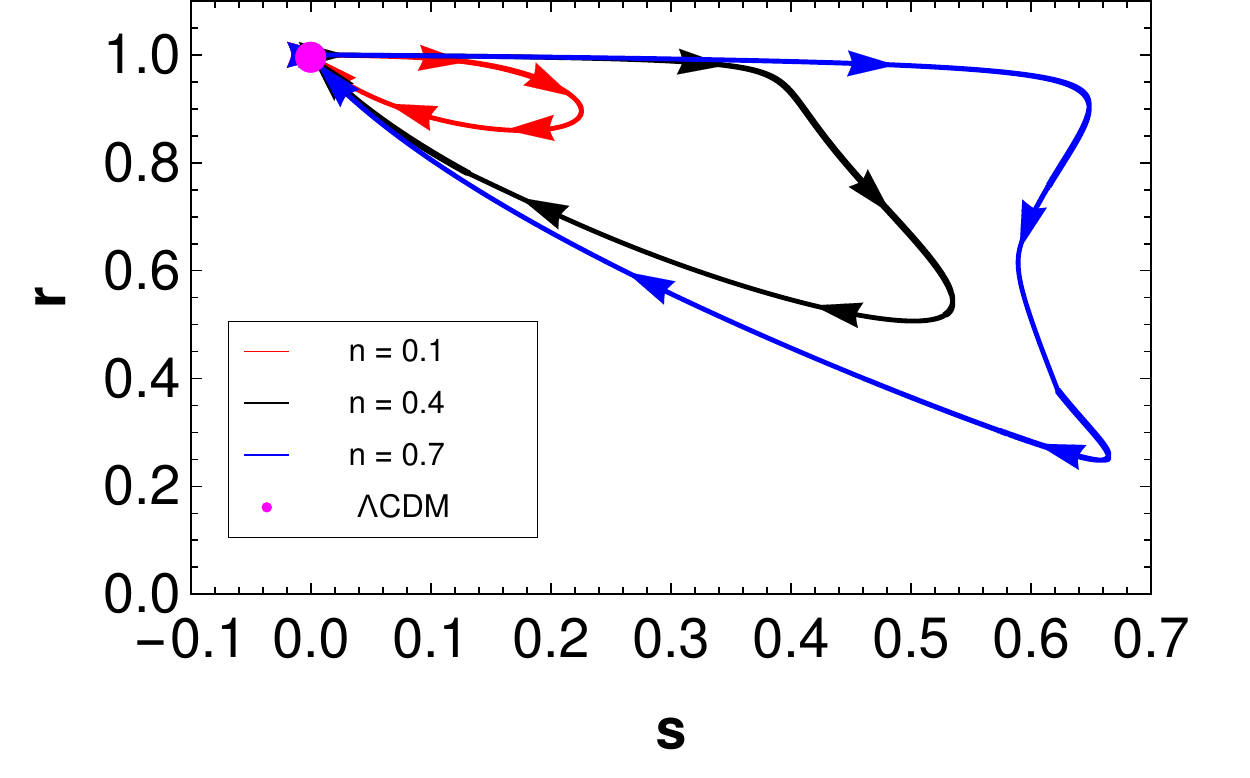}\hspace{0.5cm}
   \includegraphics[scale = 0.6]{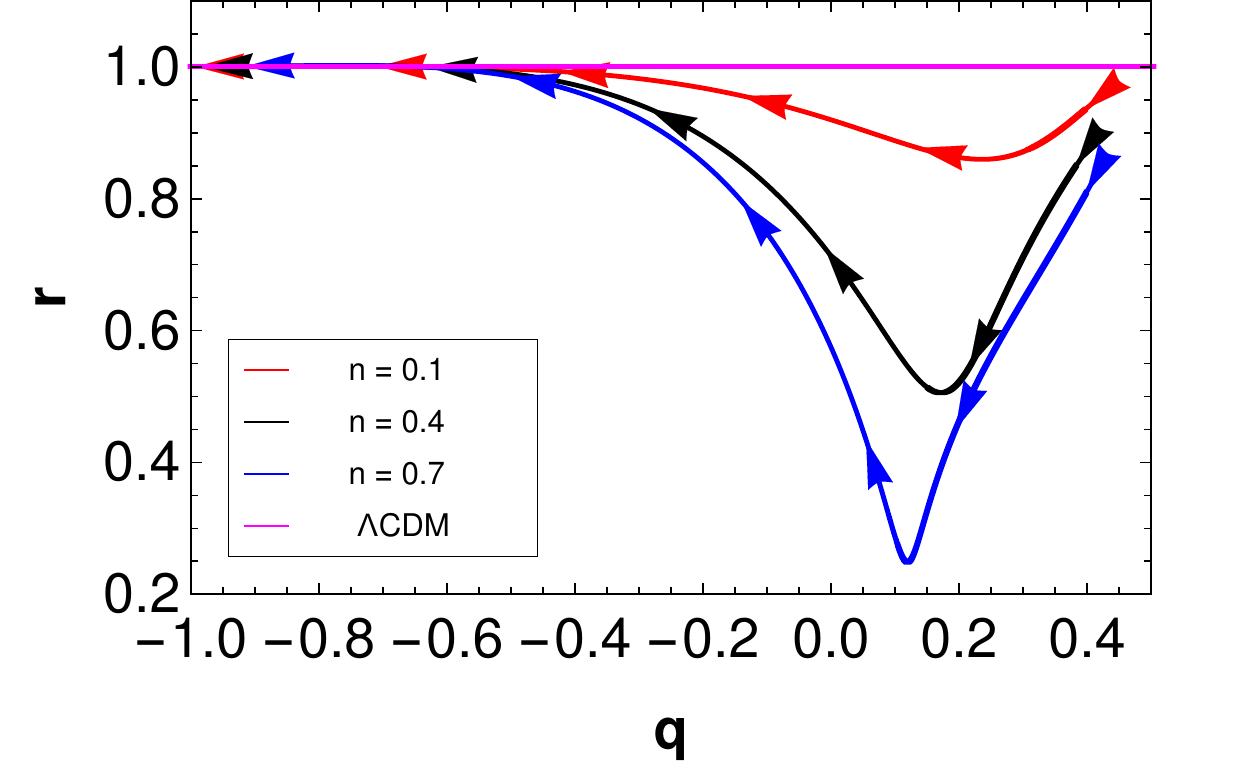}}
\vspace{-0.2cm}
\caption{Evolution of the statefinder parameter $r$ with respect to the 
parameter $s$ (left) and evolution of $r$ with the deceleration parameter $q$ 
(right) for the $f(R) = R + \alpha  H_0^{2 (1-n)} R^n$ model.}
\label{fig14}
\end{figure}

\begin{figure}[htb]
\centerline{
   \includegraphics[scale = 0.6]{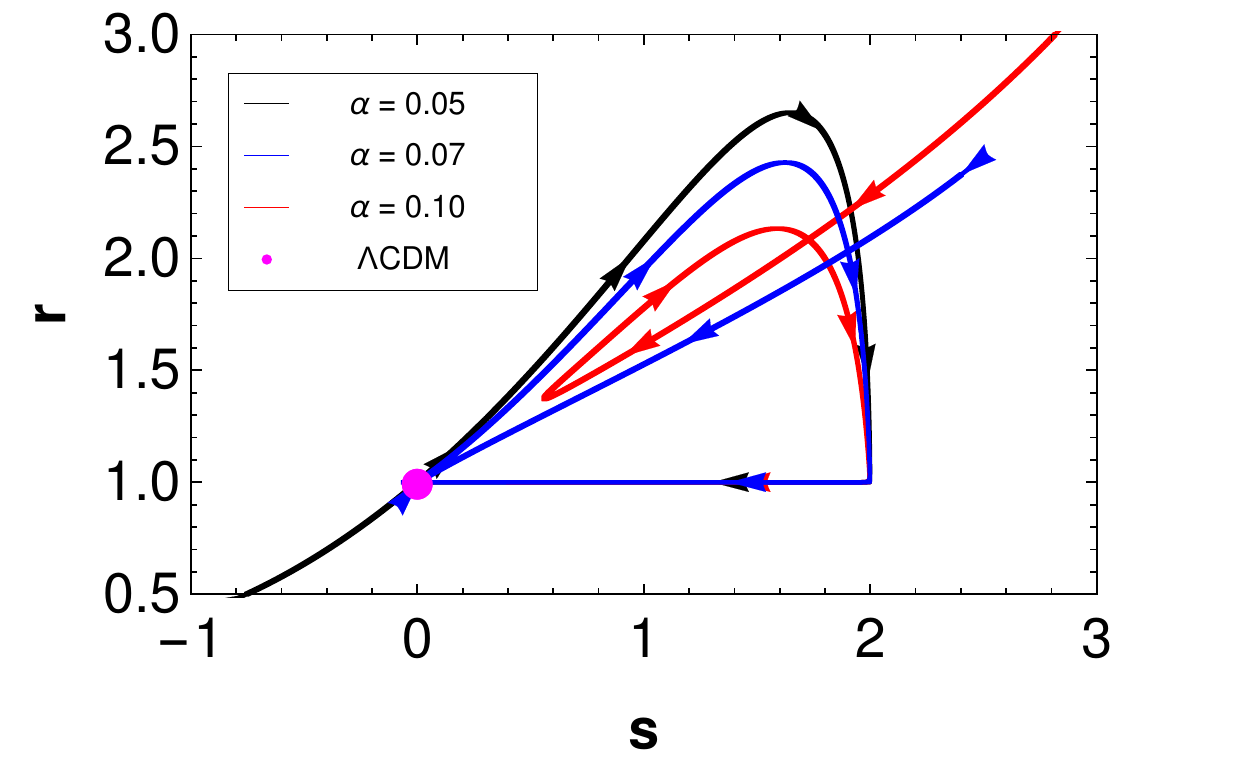}\hspace{0.5cm}
   \includegraphics[scale = 0.6]{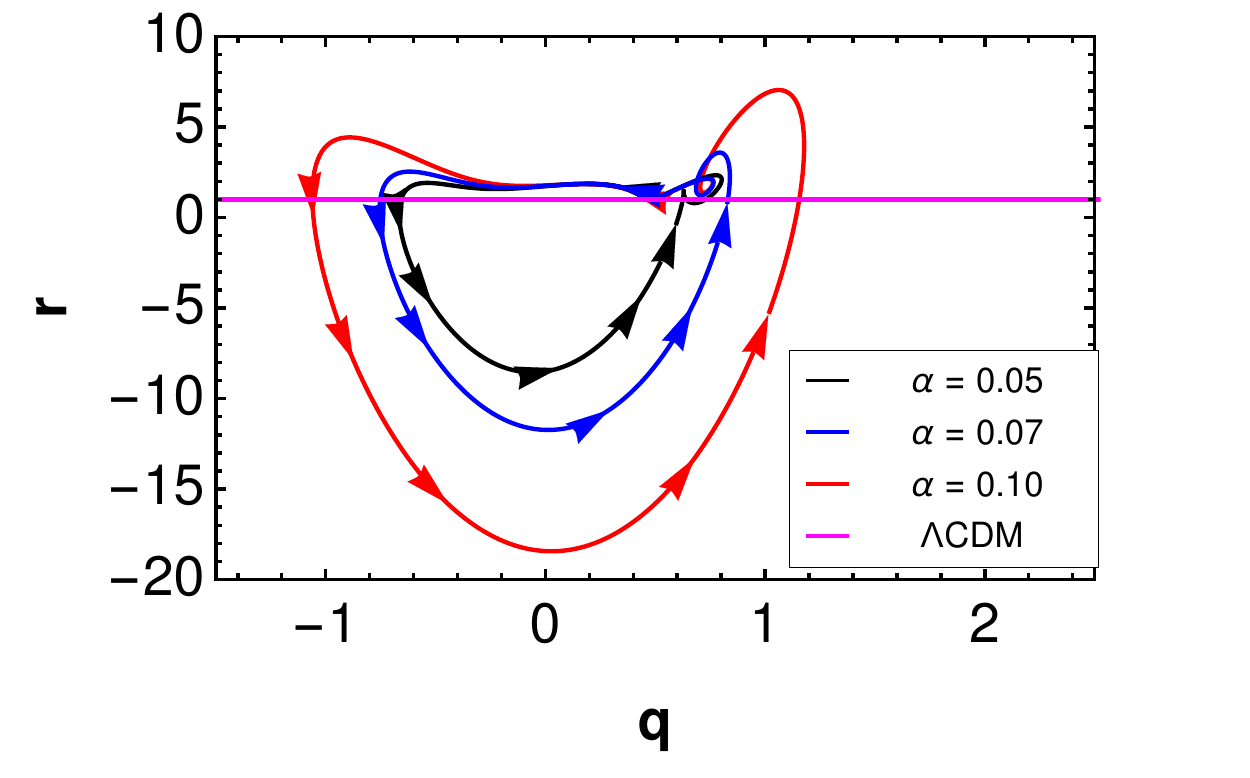}}
\vspace{-0.2cm}
\caption{Evolution of the statefinder parameter $r$ with respect to the 
parameter $s$ (left) and evolution of $r$ with the deceleration parameter $q$ 
(right) for the Gogoi-Goswami model.}
\label{fig15}
\end{figure}

We have shown the variation of the parameter $r$ with respect to $z$ for the 
power law model, $f(R) = R + \alpha  H_0^{2 (1-n)} R^n$ model and Gogoi-Goswami
model in Fig.\ \ref{fig12}. It is seen that for the power law model, the 
parameter $r$ behaves differently and can not mimic the $\Lambda$CDM model 
behaviour. Moreover, for $n=1.25$ and $n=1.38$ it crosses the $\Lambda$CDM 
model at two distinct points in the early universe and the model starts 
showing huge deviations just after or before these two points. On the other 
hand, for the case $n=1.90$, the model curve has less slope or inclination in 
comparison to the two other cases and is expected to intersect the $\Lambda$CDM
model in an earlier state of the universe. So, the model as a whole, behaves 
very differently from that of the $\Lambda$CDM model in the present or past 
universe. In case of the $f(R) = R + \alpha  H_0^{2 (1-n)} R^n$ model, we see 
that for small values of $n$ i.e.\ for $n\le0.4$, the model can not be 
efficiently differentiated from $\Lambda$CDM model in the present or near 
future. The model shows maximum deviations from the $\Lambda$CDM model in the 
near past and then again starts approaching the $\Lambda$CDM model in the early
universe. So, in the present universe or in the very early universe, the model 
can mimic the $\Lambda$CDM model and it might be difficult to differentiate it 
from the $\Lambda$CDM model in such a scenario. Similarly, in case of the 
Gogoi-Goswami model, we observe that the model shows maximum deviations from 
the $\Lambda$CDM model in the present universe and in the near future or early 
universe it shows similar behaviour as that of the $\Lambda$CDM model. So in 
the early universe, the model can not be efficiently differentiated from the 
$\Lambda$CDM. At the present scenario, deviations from the $\Lambda$CDM model 
increases from $\alpha = 0.05$ to $\alpha=0.07$.

The evolution trajectories of $(r,s)$ pair in $r-s$ plane and $r-q$ plane for 
these three models are shown in Fig.s \ref{fig13}, \ref{fig14} and \ref{fig15} 
respectively. For the power law model, for $n=1.25$ and $n=1.38$, the evolution trajectory starts from $(r,s) = (1,0)$, the statefinder pair of the 
$\Lambda$CDM model and moves in a similar path towards the present universe 
at which the deviation from each other becomes significant. However, for 
$n=1.90$ the trajectory is completely different from the previous two. Here 
the trajectory covers a different path and never moves near the point 
$(r,s) = (1,0)$. Similarly, in the $r-q$ plane trajectories for $n=1.25$ and 
$n=1.38$ follow a similar trend. For $n=1.90$, the evolution of trajectory in 
the $r-q$ plane is different from $n=1.25$ and $n=1.38$.

For the $f(R) = R + \alpha  H_0^{2 (1-n)} R^n$ model, the evolution 
trajectories of $r-s$ start from the statefinder point of the $\Lambda$CDM 
model and then covering a distinct path in the $r-s$ plane they again meet the 
statefinder point of the $\Lambda$CDM model as shown in Fig.\ \ref{fig14}. For 
$n=0.1$, the statefinder curve covers a small area in comparison to the cases 
with $n=0.4$ and $n=0.7$. The statefinders show that there are obvious 
fluctuations in the $r$ parameter and it can be a good measure to differentiate between different $n$ values as well as other models. The $r-q$ plane shows the
variation of the statefinder parameter $r$ with respect to the deceleration 
parameter $q$. Similar to the $r-s$ plane, one can see that the evolution 
trajectories in $r-q$ plane mimic the $\Lambda$CDM model in the early universe 
and towards the present and near future universe. So, before $q=0.4$ and after 
$q=-0.4$ the $r-q$ plane trajectories may not effectively differentiate the 
model from the $\Lambda$CDM model.

Similarly, for the Gogoi-Goswami model, the $r-s$ and $r-q$ evolution 
trajectories are shown in Fig.\ \ref{fig15}. It is to be mentioned here that 
for a small change in the value of $z$ the statefinder parameters especially, 
$s$ shows higher fluctuation of its value and both of them give higher values
for a given $z$ in this model in comparison to the other two models. Hence,
for this model the trajectories in the $r-s$ plane are shown only for a small 
region from around the present universe to the near future. Figure shows that 
the evolution trajectories arrive from different directions and cross the 
statefinder point of the $\Lambda$CDM model in a future point and then again 
covers a wide path to attain the point $(2,1)$, where all the curves converge 
together. From this point onwards, the trajectories move to the statefinder 
point of the $\Lambda$CDM model as a final destination. Unlike for the 
previous models, here the trajectories cover a different pattern which enables 
to differentiate it from other models in terms of statefinder parameters. The 
$r$ fluctuations in the trajectories show that the evolution of $r$ might be 
useful to differentiate the model behaviour. On the right panel of Fig.\ 
\ref{fig15}, we have plotted the statefinder parameter $r$ versus deceleration 
parameter $q$ evolution trajectories. From this plot one can see the unique 
behaviour of the model. The trajectories start moving in the positive $r$ 
side initially in the early universe. After a long journey, the trajectories 
move to the negative $r$ side and start moving in the opposite direction and 
finally at the end, they again move to the positive $r$ side and return to 
the initial point mimicking the $\Lambda$CDM model. Due to such unique 
behaviours of the model in $r-s$ and $r-q$ plane, it is possible to 
differentiate the model easily from the $\Lambda$CDM model, power law type 
model and model of the type $f(R) = R + \alpha  H_0^{2 (1-n)} R^n$.

\section{Conclusions}\label{sec07}
We have investigated the evolution of the universe with the help of  
cosmographic parameters like Hubble parameter, deceleration parameter, 
effective equation of state etc.\ as predicted by three $f(R)$ gravity models, 
viz., (i) the power law $f(R)$ gravity model $\xi R^n$, (ii) 
$f(R) = R + \alpha  H_0^{2 (1-n)} R^n$ model and (iii) the Gogoi-Goswami model 
for different values of the parameters of these models. As the Hubble 
parameter determines 
the expansion rate of the universe, we have plotted $H(z)$ versus $z$ for these 
models to check their behaviours in expansion rate at different stages of the 
universe. We have seen that for the considered values of the model parameters, 
the models show the consistency with the HKP data, SVJ05 data, SJVKS10 data and 
GCH09 data. Similarly, we have also calculated distance modulus for the models 
with the considered model parameter values sets and see that the models behave 
in a viable manner with the Union2.1 data. From these comparison plots, we 
can see that for the power law model, $n=1.38$ and $n=1.90$ show the promising 
possibility to be in the feasible range of parameter $n$. This is reflected in 
the fitting with the OHD data, which show that $n=1.40$ with $H_0 = 59.8$ has 
the best fitting results with OHD. On the other hand, the 
$f(R) = R + \alpha H_0^{2 (1-n)} R^n$ model shows a better result than the 
power law model for the considered set of parameter values. Initial study with 
the $H(z)$ versus $z$ plots and distance modulus versus $z$ plots show that 
for the parameter $n=0.1$, the model behaves in a better manner with the 
respective experimental results. Later, analysis with OHD data suggests that 
the model can show a good fitting for $n=0.1$ with $H_0=70\ \rm km/s/Mpc$ and 
$\Omega_{m0}=0.24$, which is in accordance with the previous results. The 
third model we have considered here is a very new dark energy $f(R)$ gravity 
model and the cosmological behaviour of this model has not been studied till 
now. So undoubtedly, this is the first study dealing with the cosmological 
perspectives of the model. The $H(z)$ versus $z$ plots and distance modulus 
versus $z$ plots for this model show comparatively good behaviour in 
comparison to the previous two models. It is seen from these plots that for a 
smaller value of the parameter $\alpha$ the model behaves in a better way with 
the observed data, although there is no significant difference of the results 
obtained for the allowed range of values $\alpha$. Here, we see that for 
$\alpha=0.05$, the model predictions are closer to that of the $\Lambda$CDM 
model. The OHD best fit values for the model are found to be $\alpha=0.1$, 
$\Omega_{m0}=0.30$ and $H_0=68.6\ \rm km/s/Mpc$ which are in good agreement 
with the previous results.

For more detailed information on the models, we have performed the $Om(z)$ 
test on the models. We see that the first model is significantly different
from the $\Lambda$CDM model, whereas the second and the third model can 
mimic the $\Lambda$CDM model in the early universe. However, as a whole, for 
the second model i.e.\ the $f(R) = R + \alpha  H_0^{2 (1-n)} R^n$ model, the 
$Om(z)$ function gives higher values than that of the $\Lambda$CDM model. But 
in the Gogoi-Goswami model, the $Om(z)$ function initially starts with a higher 
value than that of the $\Lambda$CDM model, then drops below the 
$Om(z)_{\Lambda CDM}$ and finally starts approaching the $\Lambda$CDM model 
in the early universe. This behaviour suggests that the Gogoi-Goswami model 
behaves differently than the first and the second model. Similar results are 
obtained in the statefinder analysis also. We have seen that the evolution 
trajectories of the statefinder parameters behave uniquely in the 
Gogoi-Goswami model and the model can be easily differentiated from the 
previous two models with the help of statefinder parameters.
Thus, it is seen that although the $\Lambda$CDM can be reconstructed by a huge 
class of extended gravity models with the help of model parameters, usually 
the models can be differentiated from each other by these two most effective
diagnostic tests. Also, one may consider the Bayesian method for the Bayesian 
information criteria for a better comparison of the dark energy models 
including modified gravity models \cite{bayesian}. We leave this method of 
study as a future scope of our works. Moreover, to comment more about the 
viabilities of such models, one may consider the studies of inflationary 
epoch, large scale structure etc.\ from such models \cite{inflationary}. So we 
believe that another study in this perspective will shed more light to the 
properties and behaviour of the new model.

Thus from our study, we conclude that the Palatini Gogoi-Goswami model can be 
a good alternative to the $\Lambda$CDM model and can result in more interesting 
outcomes in predicting the fate and evolution of the universe. In the near future 
this model can be used in the metric formalism as well as in the Palatini 
formalism to test it in the context of various observational astrophysical 
and cosmological data.

\end{document}